\documentclass[a4paper,fleqn,usenatbib]{mnras}
\usepackage{Times}
\usepackage{psfrag}
\usepackage{pspicture}
\usepackage{graphicx}
\usepackage{amsmath}
\usepackage{textcomp}
\usepackage[T1]{fontenc}
\usepackage{ae,aecompl}

\newcommand{\hb}{H$\beta$ }
\newcommand{\z}{\textit{z}}
\newcommand{\til}{$\sim$}

\title[The most luminous H$\alpha$ emitters at $z\sim0.8-2.2$]{The most luminous H$\bf \alpha$ emitters at $\bf z\sim0.8-2.23$ from HiZELS: evolution of AGN and star-forming galaxies\thanks{This research is based on data gathered at the ESO New Technology Telescope under programs 087.A-0337 and 089.A-0965, Telescopio Nazionale Galileo, with time awarded through OPTICON programs 2011A/026 and 2012A020 and the William Herschel Telescope under program W12BN007.} }

\author[D. Sobral et al.]{David Sobral$^{1,2,3,4}$\thanks{VENI/FCT IF Fellow. E-mail: sobral@strw.leidenuniv.nl},  Saul A. Kohn$^{3,5}$, Philip N. Best$^{6}$, Ian Smail$^{7}$, Chris M. Harrison$^{7}$ \newauthor John Stott$^{7,8}$, Jo\~{a}o Calhau$^{1,2,4}$, Jorryt Matthee$^{3}$ \\
$^{1}$ Instituto de Astrof\'{\i}sica e Ci\^{e}ncias do Espa\c{c}o, Universidade de Lisboa, OAL, Tapada da Ajuda, PT1349-018 Lisboa, Portugal \\
$^{2}$ Departamento de F\'{i}sica, Faculdade de Ci\^{e}ncias, Universidade de Lisboa, Edif\'{i}cio C8, Campo Grande, PT1749-016 Lisbon, Portugal \\
$^{3}$ Leiden Observatory, Leiden University, P.O.\ Box 9513, NL-2300 RA Leiden, The Netherlands\\
$^{4}$ Department of Physics, Lancaster University, Lancaster, LA1 4YB, UK \\
$^{5}$ Department of Physics and Astronomy, University of Pennsylvania, Philadelphia, PA, 19104, USA \\
$^{6}$ SUPA, Institute for Astronomy, Royal Observatory, Blackford Hill, Edinburgh, EH9 3HJ, UK\\
$^{7}$ Centre for Extragalactic Astronomy, Department of Physics, Durham University, South Road, Durham DH1 3LE UK \\
$^{8}$ Sub-department of Astrophysics, Dpt. of Physics, University of Oxford, Denys Wilkinson Bld., Keble Road, Oxford OX1 3RH, UK
}

\begin{document}

\date{Accepted 2016 January 04. Received 2016 January 04; in original form 2015 October 10}

\pagerange{\pageref{firstpage}--\pageref{lastpage}} \pubyear{2016}
\maketitle

\label{firstpage}
\begin{abstract}
We use new near-infrared spectroscopic observations to investigate the nature and evolution of the most luminous H$\alpha$ emitters at $z\sim0.8-2.23$, which evolve strongly in number density over this period, and compare them to more typical H$\alpha$ emitters. We study 59 luminous H$\alpha$ emitters with $L_{\rm H\alpha}>$\,$L_{\rm H\alpha}^*$, roughly equally split per redshift slice at $z\sim0.8$, $1.47$ and $2.23$ from the HiZELS and CF-HiZELS surveys. We find that, overall, $30\pm8$\,\% are AGN ($80\pm30$\% of these AGN are broad-line AGN, BL-AGN), and we find little to no evolution in the AGN fraction with redshift, within the errors. However, the AGN fraction increases strongly with H$\alpha$ luminosity and correlates best with $L_{\rm H\alpha}$/$L_{\rm H\alpha}^*(z)$. While $L_{\rm H\alpha}$\,$\le$\,$L_{\rm H\alpha}^*(z)$ H$\alpha$ emitters are largely dominated by star-forming galaxies ($>80$\,\%), the most luminous H$\alpha$ emitters ($L_{\rm H\alpha}>10L_{\rm H\alpha}^*(z)$) at any cosmic time are essentially all BL-AGN. Using our AGN-decontaminated sample of luminous star-forming galaxies, and integrating down to a fixed H$\alpha$ luminosity, we find a factor of $\sim1300\times$ evolution in the star formation rate density from $z=0$ to $z=2.23$. This is much stronger than the evolution from typical H$\alpha$ star-forming galaxies and in line with the evolution seen for constant luminosity cuts used to select `Ultra-Luminous' Infrared Galaxies and/or sub-millimetre galaxies. By taking into account the evolution in the typical H$\alpha$ luminosity, we show that the most strongly star-forming H$\alpha$-selected galaxies at any epoch ($L_{\rm H\alpha}>L^*_{\rm H\alpha}(z)$) contribute the same fractional amount of $\approx15$\% to the total star-formation rate density, at least up to $z=2.23$.
\end{abstract}


\section{Introduction}
\label{sec:intro}

Surveys show that the peak of the star-formation history \citep[e.g.][]{Lilly.96, Hopkins.06, Reddy.08, Sobral.13,Swinbank2014} and active galactic nuclei \citep[AGNs, e.g.][]{Wolf.03, Ackermann.11} activity lies within the redshift interval $z=1-3$, although there are suggestions that the peak of star formation activity may occur earlier \citep[$z\sim2-2.5$; e.g][]{Karim.11,Sobral.13}, than that of AGN \citep[$z\sim1-2$; e.g.][]{Aird2010,Ueda2014}. There is also evidence that the strong evolution in star formation activity has happened for galaxies at all masses \citep[e.g.][]{Sobral.14,Drake2015} and in all environments \citep[e.g.][]{Koyama.13}.

H$\alpha$ is an excellent instantaneous tracer of star formation activity \citep[e.g.][]{Kennicutt.98}. By using the narrow-band technique \citep[see also grism surveys, e.g.][]{Colbert.13} on very wide-field NIR detectors, H$\alpha$ can be used to conduct very large, sensitive surveys up to $z\sim2.5$ \citep[e.g.][]{Kurk2004,Geach.08,Lee2012,Koyama.13,Sobral.13,Stroe.Sobral15}. Measuring the evolution of the H$\alpha$ luminosity function (LF) is one of the main goals of High-$z$ Emission Line Survey \citep[HiZELS,][]{Geach.08, Sobral.09a, Sobral.12, Sobral.13, Best2013}. Using the H$\alpha$ emission line as a star-formation indicator allows the use of the same robust, well-calibrated and sensitive indicator over $\sim11$\,Gyrs of cosmic time. Several studies have now explored this unique potential, both from the ground and from space, to unveil the evolution of morphologies, dynamics and metallicities of star-forming galaxies \citep[e.g.][]{Fumagalli.12, Livermore.12, Swinbank.12, Colbert.13, Dominguez.13, Koyama.13, Price.13, Sobral.13b, Stott13a, Stott13b, Tadaki.13, An.14,Darvish2014,Sobral2015S}.
  
These studies show that in the last 11 Gyrs since $z\sim2.5$, the space density of luminous H$\alpha$ emitters ($L_{\rm H\alpha}>10^{42}$\,erg\,s$^{-1}$) has dropped by several orders of magnitude \citep[e.g.][]{Geach.08,Lee2012,Sobral.13,Colbert.13,Sobral15,Stroe.Sobral15}. \cite{Sobral.13} find that the strong evolution in the H$\alpha$ luminosity function (LF) is best described by the evolution of the typical H$\alpha$ luminosity ($L_{\rm H\alpha}^*$) with redshift (although $\Phi^*$ is also shown to evolve). In practice, studies find an order of magnitude increase in the characteristic $L_{\rm H\alpha}^*$ or the knee of the star formation rate function, SFR$^*$ from $z\sim0$ to $z\sim2.2$ \citep{Geach.08,Sobral.09a,Hayes2010,Lee2012,Sobral.13,Sobral.14}. Similar evolution is found for other nebular lines such as [O{\sc ii}] and H$\beta$+[O{\sc iii}] \citep[e.g.][]{Khostovan2015}. Interestingly, the most significant changes within the H$\alpha$ population at the peak of star formation history seem to be driven by or linked to this strong luminosity evolution of H$\alpha$ emitters \citep[c.f.][]{Sobral.09a, Sobral.12, Sobral.13}. In fact, when one normalises H$\alpha$ luminosities by L$_{\rm H\alpha}^*(z)$, taking into account its evolution with redshift, many of the relations with luminosity, that would be found to evolve with redshift, fall back, to first order, to an almost-invariant relation that does not depend on cosmic time \citep[see e.g. clustering and dust extinction studies; ][]{Sobral.10, Sobral.12,Stott13b}. Clustering studies have shown that $\sim$L$_{\rm H\alpha}^*$ galaxies seem to have been hosted by Milky Way mass haloes ($\sim10^{12}$\,M$_{\odot}$) at least since $z\sim2.2$ \citep{Sobral.10, Geach.12,Stroe.Sobral15}, while $\gg L_{\rm H\alpha}^*$ emitters seem to reside in 10$^{13}$M$_{\odot}$ or higher mass dark matter haloes. This may be important if we are to understand the processes that may quench the most massive galaxies. It is therefore essential to understand the nature of such luminous H$\alpha$ sources and whether they host active galactic nuclei (AGN).

While there are many ways to identify AGNs within a sample of emission line galaxies, including the use of X-rays, radio, and mid-infrared \citep[e.g.][]{Lacy.04,Lacy.07,Garn.10,Stern.12,BrandtAlex15}, rest-frame optical spectroscopy is still one of the most robust ways to identify AGN. The identification of AGN is particularly simple and straightforward in the presence of luminous broad Balmer lines (typically FHWM well in excess of $1000$\,km\,s$^{-1}$), which indicate AGN. For narrow-line emission line sources, well-chosen emission line ratios are the most robust way to identify any AGN. \cite{Baldwin.81} were among the first to recognise the importance of emission line ratios for distinguishing star-forming galaxies (SFGs) from AGNs. Their diagnostic (the `BPT diagram') was based on using the relative line intensities in order to reveal the dominant excitation mechanism that operates upon the line-emitting gas: photoionisation by OB stars (in star-forming galaxies, SFGs) or by a power-law continuum (AGNs). More recently, \cite{Kewley.13} argue that the BPT calibration needs to be adjusted to account for the redshift evolution in the interstellar medium conditions and radiation field, which is observed in galaxies across cosmic time.

Many studies have sought to use the BPT and similar emission-line diagnostics to reveal the nature of low to intermediate redshift galaxies \citep[e.g.][]{Brinchmann.04,Obric.06,LaMassa.12}. \cite{Smolcic.06} found a tight correlation between the rest-frame colours of emission line galaxies and their position on the BPT diagram. Other studies have used spectral energy distribution (SED) template-fitting to separate AGNs and SFGs within large samples \citep[e.g.][]{Fu.10,Kirkpatrick.12} by exploring Spitzer spectroscopy and imaging, combined with deep {\it Herschel} data. Such studies find evidence for a co-evolution scenario (at least since $z\sim1$), in which a period of intense accretion onto the central black hole of a galaxy may coincide with starburst episode, but over different timescales \citep[e.g.][]{Fu.10}. Other studies have focused on Lyman-break galaxies \citep[e.g.][]{Schenker.13} and X-ray selected sources \citep[e.g.][]{Trump.13}. Indeed, with instruments such as KMOS \citep[for early science results see e.g.][]{Sharples.06,Sobral.13b,Stott2014,Wuyts.14} and MOSFIRE \citep[e.g.][]{McLean.08,Kriek15} now fully operational, many more similar and larger studies will be possible, but those will be mostly focusing on $\le L_{\rm H\alpha}^*$ galaxies. Despite the interest in, and importance of highly luminous emitters in the high-redshift H$\alpha$ luminosity function (LF), little is known about them due to the difficulty of consistently selecting targets and following them up spectroscopically.

In this paper, we present near-IR spectroscopic observations, and subsequent analyses, of the most luminous H$\alpha$ emitting galaxies in HiZELS and CF-HIZELS \citep{Sobral.13,Sobral15}: $>L_{\rm H\alpha}^*$ H$\alpha$ emitters. Our goal is to unveil the nature of such luminous H$\alpha$ emitters and to investigate their potential evolution across cosmic time. The paper is organised as follows. In \S\ref{section:SampleObsRed} we describe the sample, observations and data reduction. \S\ref{section:MethMed} presents the redshift distributions, explains the spectral line measurements and presents the analysis. We present the results and discussion in \S\ref{sec:Res}, and unveil the nature and evolution of luminous H$\alpha$ emitters across cosmic time. We also present an AGN-decontaminated SFR-history of the Universe over the past $\sim$11\,Gyr. We summarise our findings and conclude in \S\ref{section:conc}. We use AB magnitudes, a Chabrier IMF and assume a cosmology with H$_{0}$=70kms$^{-1}$Mpc$^{-1}$, $\Omega_{M}$=0.3 and $\Omega_{\Lambda}$=0.7.

\section{Observations and Data Reduction} \label{section:SampleObsRed}

%
%
\begin{table}
\centering
\caption{A summary of the fields used in this study, the area covered with narrow-band imaging, and the number of luminous H$\alpha$ emitters targeted in each of the fields.}
\begin{tabular}{cccc}
\hline
Narrow-band	&	Field  & Area		&	\# Targets 	 \\	
 ($z$, redshift) & & (deg$^2$) & \\
\hline						
NBJ 	 & COSMOS & 0.8 & 7   \\
 ($z\sim0.8$) 	 & SA22 & 10 & 11   \\
	  & UDS & 0.6 & 5   \\
\hline	
NBH	 & Bo{\"o}tes & 0.8 & 6  \\
 ($z\sim1.47$) 	 & COSMOS & 1.6 & 7  \\
	 & SA22 & 0.8 & 10   \\
	  & UDS & 0.8 & 12   \\
\hline	
NBK	 & COSMOS & 1.6 & 9  \\
 ($z\sim2.23$)  	 & SA22 & 0.8 & 6   \\
	  & UDS & 0.8 & 6   \\
\hline
\end{tabular}
\label{fields_numbers}
\end{table}

\subsection{Sample Selection: HiZELS H$\alpha$ luminous emitters}

We selected H$\alpha$ luminous sources likely to be at $z=0.84$, 1.47 and 2.23 from HiZELS \citep{Best2013,Sobral.09a, Sobral.12, Sobral.13} and $z=0.81$ from the CF-HiZELS survey \citep[][]{Sobral.13b,Sobral15,Matthee2014}. HiZELS used the Wide Field Camera (WFCAM) on the United Kingdom Infrared Telescope (UKIRT), to select emission line galaxies at various redshifts using specially designed narrow-band filters in the $J$ (NB$_{\rm J}$) and $H$ bands (NB$_{\rm H}$), along with the H$_2$S(1) filter in the $K$ band (NB$_{\rm K}$). HiZELS surveyed $\sim0.8$\,deg$^2$ contiguous regions in UKIDSS \citep{Lawrence.12} UDS and SA22 fields, the Bo{\"o}tes field \citep[e.g.][]{Brand.06} and $\sim1.6$\,deg$^2$ in the COSMOS field \citep{Scoville.07} field. CF-HiZELS used WIRCAM on CFHT to obtain a $\sim$10\,deg$^2$ narrow-band survey in the $J$ band over the SA22 field. The addition of the $z=0.81$ CF-HiZELS sample allows us to select luminous emitters at $z\sim0.8$ by probing a volume much more comparable to that of $z\sim1.47$ and $z\sim2.23$ HiZELS studies. For the rest of the paper, we will refer to the sample at $z=0.81$ and $0.84$ as our $z\sim0.8$ sample. For SA22 and Bootes (where photometric data do not reach the excellency level of COSMOS and UDS), and in order to assure a high completeness, we opt to follow up all of the brightest line emitters selected from each of the narrow-band filters. For UDS and COSMOS, we use the sample presented by \cite{Sobral.13}. Briefly, the sample of H$\alpha$ emitters is selected (isolating H$\alpha$ emitters from other higher and lower redshift line emitters) using i) spectroscopic redshift confirmation with other emission lines, ii) photometric redshifts and iii) colour-colour selections to exclude non-H$\alpha$ emitters. We refer the reader to \cite{Sobral.13} for further details. 

Our choice of flux cuts was motivated by the need to consistently trace luminous H$\alpha$ emitters across redshifts. To this end, we took into account the increase in the knee of H$\alpha$ LF (L$_{\rm H\alpha}^*$) with redshift \citep{Sobral.13}\footnote{Note that the equation presented in \cite{Sobral.13} was derived assuming A$_{\rm H\alpha}=1$\,mag, and thus to correct back to the observed L$^*$ one needs to subtract 0.4\,dex. The version presented here is for observed H$\alpha$ luminosities.}: 

\begin{equation} \label{eq:Lhastar}
\log L_{\rm H\alpha}^* = 0.45z + 41.47.
\end{equation}
We then selected sources which had luminosities corresponding to $\geq1.0$~L$_{\rm H\alpha}^*$(\z) and reaching up to $\sim 50$\,$L_{\rm H\alpha}^*$(\z), with number densities in the range 10$^{-3.2}$\,Mpc$^{-3}$ to 10$^{-6}$\,Mpc$^{-3}$. Our luminosity limits roughly correspond to (observed) fluxes greater than $\sim3$, $\sim2$ and $\sim1.5\times10^{-16}$\,erg~s$^{-1}$~cm$^{-2}$ for $z \sim$~0.8,~1.47 and 2.23, respectively.
Our targets are distributed across the HiZELS fields: the UKIDSS UDS and SA22 fields, the COSMOS field and the Bo{\"o}tes field (see Table \ref{fields_numbers}). For full details on the catalogues, see \cite{Sobral.13} and \cite{Sobral15}.

%
%
%
\begin{table*}
\caption{Observing log for the different instruments and grisms used in this study. Each one was capable of observing the H$\alpha$, [N{\sc ii}], [O{\sc iii}] and H$\beta$ lines, with the exception of the Blue ($YJH$) grism on SofI, which does not cover [O{\sc iii}] and H$\beta$ for z$\sim$0.81 targets. The NB$_J$ sample targets H$\alpha$ emitters at $z\sim0.8$, the NB$_H$ sample targets H$\alpha$ emitters at $z=1.47$, and the NB$_K$ sample targets H$\alpha$ emitters at $z=2.23$. Pixel scale is given as observed and in rest-frame (R.F.).}
\label{OBSERVATIONS}
\begin{tabular}{cccccccc}
\hline
Instrument & Grism 	&	 $\lambda$ coverage 	&	Sample  & \# Sources	& Pix. Scale (R.F.) & Dates Observed & Typical Seeing \\
    &    	&	(\AA, observed) 	&	 & (H$\alpha$)  & (\AA /pixel)	& & ($''$)\\
\hline 							  
WHT/LIRIS & lr\_zj 	      &     8870--15310  &	NB$_J$  & 8    &	6 (3.3)  & 16-19 Jan 2013 & 0.7  \\  
NTT/SofI & Blue  	&	 9500--16400 	&	NB$_J$ & 9	 &	7 (3.8)  & 18-21 Sep 2012 & 0.7	\\
NTT/SofI & Blue 	&	 9500--16400 	&	NB$_H$ & 20	 &	7 (2.8) & 23-25 Sep 2011 & 0.6	\\
TNG/NICS & $JH$ 	    &    11500--17500  &	NB$_H$  & 8	&	7 (2.8)  &  26 Apr 2011, 1-4 Apr 2012 & 1.5\\
WHT/LIRIS & lr\_hk  	&	 13880--24190  	&	NB$_K$ &  8 &	10 (3.1	) &16-19 Jan 2013 & 0.7\\
NTT/SofI & Red	&	 15300--25200 	&	NB$_K$ &  6	&	10 (3.1) & 18-21 Sep 2012 & 0.8\\
\hline 
\end{tabular}
\end{table*}

\subsubsection{NB$_J$ sample (H$\alpha$ $z\sim0.8$)} 

We selected 23 candidate line emitters with narrow-band $J$ (NB$_J$) estimated line fluxes higher than $3.0\times10^{-16}$\,erg\,s$^{-1}$\,cm$^{-2}$ (average flux of $7.3\times10^{-16}$\,erg\,s$^{-1}$\,cm$^{-2}$): 11 from SA22, 5 from UDS and 7 from the COSMOS field.

\subsubsection{NB$_H$ sample (H$\alpha$ $z=1.47$)} 

We selected 35 candidate line emitters (likely H$\alpha$ emitters at $z=1.47$) with the highest narrow-band (NB$_H$) fluxes, $>2.0\times10^{-16}$\,erg\,s$^{-1}$\,cm$^{-2}$ (average flux of $1.1\times10^{-15}$\,erg\,s$^{-1}$\,cm$^{-2}$). From these, 12 sources are from the UDS field, 10 from SA22, 7 sources are found in the COSMOS field, and the remaining 6 sources are from the Bo{\"o}tes field (without a colour-colour or photometric redshift pre-selection, thus more likely to have contaminants).

\subsubsection{NB$_K$ sample (H$\alpha$ $z=2.23$)} 

For our sample at $z=2.23$, we select a total of 21 sources: 9 sources from COSMOS, 6 from UDS and 6 sources from SA22 (without a colour-colour or photometric redshift pre-selection). We select them for being NB$_K$ emitters with NB estimated line fluxes $>1.5\times10^{-16}$\,erg\,s$^{-1}$\,cm$^{-2}$ (average flux of $5.7\times10^{-16}$\,erg\,s$^{-1}$\,cm$^{-2}$).

\subsubsection{Comparison sample: NB$_J$ and NB$_H$ follow-up with FMOS}

In order to explore a wider parameter space, and to be able to compare our luminous H$\alpha$ emitters with those which are much more typical at their redshifts, we use a comparable spectroscopic sample of lower luminosity H$\alpha$ emitters at $z=0.84$ and $z=1.47$ from \cite{Stott13b}, observed with FMOS on the 8-m Subaru telescope. Because the sample is the result of follow-up of candidate H$\alpha$ emitters from exactly the same parent samples as we are using here, it is an ideal sample to compare our results with more ``typical" sources.

\subsection{Spectroscopic Observations}

\begin{figure*}
\includegraphics[width=13.5cm]{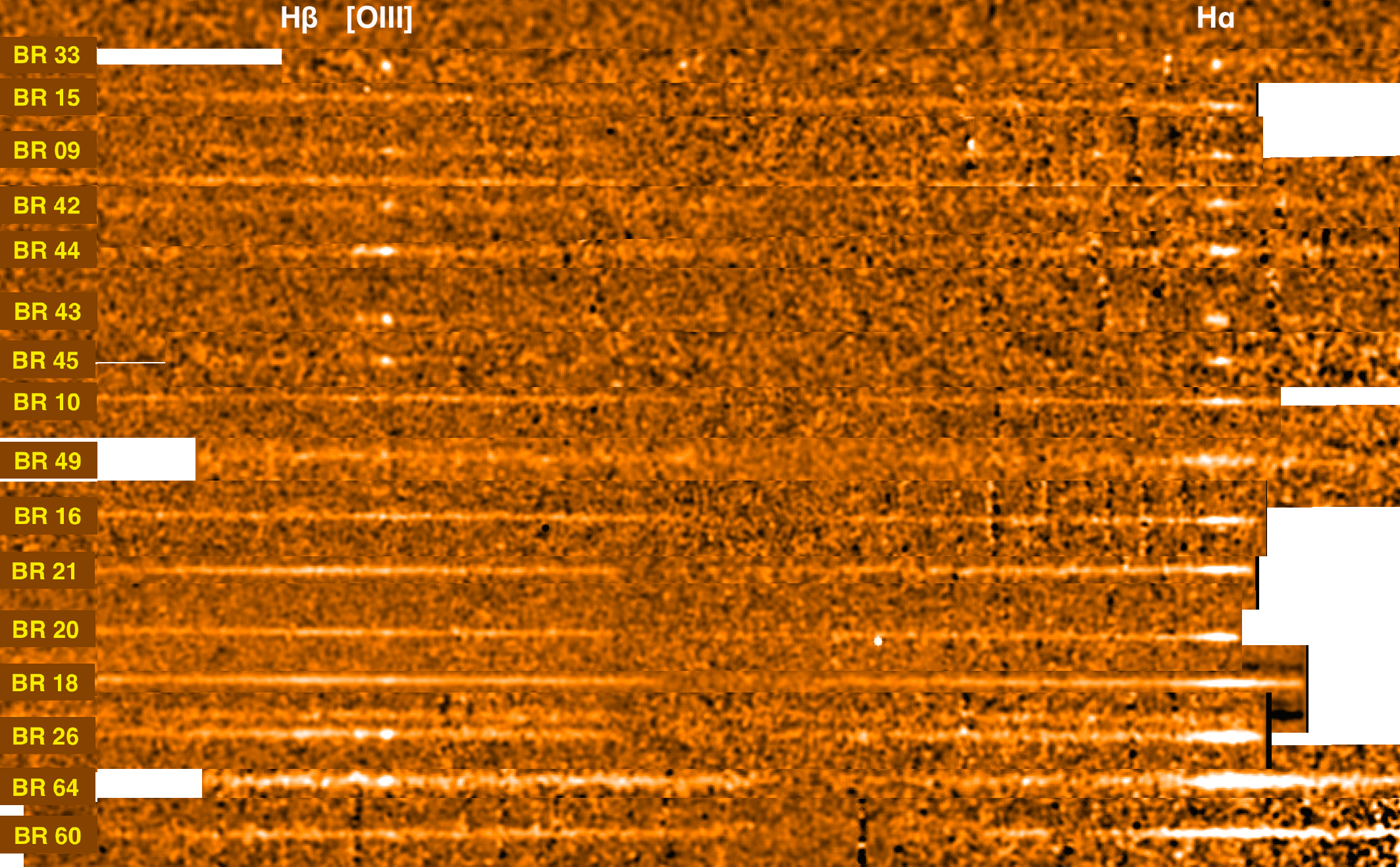}
\caption{Example 2D reduced spectra from WHT/LIRIS (top and bottom sources; BR-33 and BR-60) and NTT/SofI (remaining sources) at $z=0.8$, $z=1.47$ and $z=2.23$. Note that white spaces/regions are due to the slightly different (rest-frame) spectral coverage. We show examples which, from top to bottom, represent an increase in H$\alpha$ flux and H$\alpha$ FHWM. At the highest fluxes, our sample is dominated by broad-line AGN. Note that there is a range in wavelength which corresponds to the region between either $Y$ and $J$, $J$ and $H$ or $H$ and $K$, where the atmospheric transmission is extremely low, and thus the apparent drop in the continuum, for sources where the continuum is detected. There are no emission lines in that region, and thus we neglect it for the analysis. Our broadest H$\alpha$ line emitters are BR-64 and BR-60: these indicate high accretion speeds present within the galactic nuclei and/or outflows. BR-64, with $z$=2.197$\pm$0.001, presents FWHM$_{\rm H\alpha}$=11500$\pm$700 km\,s$^{-1}$, while BR-60, $z$=2.207$\pm$0.001, has FWHM$_{\rm H\alpha}$=10500$\pm$1100 km\,s$^{-1}$.}
\label{2D_SPECTRA}
\end{figure*}

We observed our samples of luminous line emitter candidates in the near-IR, in order to probe the rest-frame optical and recover, with a single spectrum, H$\beta$, [O{\sc iii}], H$\alpha$ and [N{\sc ii}] -- see Figure \ref{2D_SPECTRA}. In order to achieve our goals, we used NTT/SofI, WHT/LIRIS and TNG/NICS (see Table \ref{OBSERVATIONS}). The details of our observations using each instrument are discussed next, while Figure~\ref{2D_SPECTRA} shows examples of spectra gathered using the different instruments and at the different redshifts. Typical total exposure times per source were very modest: $\sim$3\,ks\,pix$^{-1}$, but ranged from 1.8\,ks\,pix$^{-1}$ for the brightest sources to 8\,ks\,pix$^{-1}$ for the sources with the faintest observed flux.

%
%
%
%

\subsubsection{NTT/SofI: NB$_J$, NB$_H$ and NB$_K$ samples}

We used SofI \citep[Son of ISAAC;][]{Moorwood.98} on the ESO NTT in La Silla over 23-25 September 2011 and 18-21 September 2012 (see Table \ref{OBSERVATIONS}). We obtained spectra of sources selected from SA22 and UDS. During the 2011 run we used the 1\arcsec slit and the blue grism with $R\sim1000$ (9500--16400\,\AA, corresponding to the rest-frame range 3900--6700\,\AA \ for sources at $z\sim1.47$), which allowed simultaneous $YJH$ coverage. In 2012 the 1\arcsec slit and the blue grism (corresponding to a rest-frame range 5300--9000\,\AA \ for objects at $z\sim0.8$), and the 1\,\arcsec slit with the red grism with $R\sim1000$ (15300--25200\,\AA, corresponding to rest-frame range 4700--7800\,\AA \ for objects at $z\sim2.23$) were used. All observations were conducted under clear conditions.

Individual exposures were 200\,s in the instrument's non-destructive mode. We applied offsets along the slit for different exposures of the same target ($\sim30''$ on average), which were further jittered with smaller offsets ($\sim1-3''$) in an ABBAAB sequence for optimal sky subtraction and badpixel removal. Dome flats and dark and arc frames were taken at the beginning of each night. Telluric stars were observed 2-3 times per night at the corresponding air masses and positions to the targets. Telluric stars were reduced by following the same procedure as the science targets, and then used to calibrate the science target spectra. Three targets were acquired directly (centred on the slit directly, as they were bright enough in the continuum). For the other targets, we acquired a nearby bright source and rotated the instrument, so that both the bright source and our science target were on the slit at all times. This not only allowed us to quickly acquire and assure that the science target did not move out of the slit.

Total exposure times varied between 2.7\,ks for the most luminous sources and 6\,ks for the faintest ones. In our Sep 2011 run the seeing varied between 0.5\arcsec~ and 0.8\arcsec~ with a median of 0.6\arcsec. Seeing was similar for the 2012 run, only slightly higher, varying from 0.6\arcsec~ to 0.9\arcsec, but with an average of 0.7$''$. During our 2011 run (targeting our NB$_H$ sample), we were able to confirm 20 H$\alpha$ emitters at $z=1.47$, with a high fraction of broad-line H$\alpha$ emitters. For our 2012 run, targeting our NB$_J$ and NB$_K$ samples, we confirmed 9 H$\alpha$ emitters at $z\sim0.8$ and 6 at $z=2.23$.

\subsubsection{TNG/NICS: NB$_H$ sample} \label{subsub:tng}

We used the NICS (Near-Infrared Camera and Spectrometer) instrument \citep{Baffa.01} with the $JH$ grism ($R\sim500$) and the 1\arcsec slit to observe NB$_H$ candidate line emitters. This instrumental set up allowed us to probe 11500--17500\,\AA, allowing us to target the rest-frame range \til4700--7100\AA \ for sources at \z\til1.47 (NB$_H$ selected), which were the sole aim of the TNG runs. We used TNG/NICS to observe our targets selected from the COSMOS and Bo{\"o}tes fields on the 26th April 2011, and the 1st, 2nd and 4th April 2012. During both runs the seeing was 1-2.5\arcsec, and thus significantly worse than that for e.g. the NTT runs. Dark frames, flats and arcs were obtained at the beginning of the night. During the 2011 run we observed two targets, one in COSMOS and one in Bo{\"o}tes, which were acquired directly. We observed one telluric star after observing one of the targets and before moving to the next. During the 2012 run, targets were observed by first acquiring a nearby bright source and then rotating the instrument to align the slit with the bright source and the target. Telluric stars were taken at the beginning, middle, and towards the end of each night (so 3 telluric stars were available for calibration), taken from fields near those under observation at the time. We used individual exposure times of 300\,s. In total, using NICS, we were able to confirm 8 H$\alpha$ emitters at $z\sim1.47$ (from our NB$_H$ sample).

\subsubsection{WHT/LIRIS: NB$_J$ and NB$_K$ samples}

We used LIRIS \citep[Long-slit Intermediate Resolution Infrared Spectrograph;][]{Manchado.98} on the WHT to obtain spectra for NB$_J$ and NB$_K$ sources selected from the COSMOS and the UDS fields with the 1\arcsec slit. Over 16-19 January 2013 we obtained spectra of 23 targets in the $HK$ (probing 13880--24190\,\AA, rest-frame 4300--7500\,\AA \ for sources at $z\sim2.23$) and ZJ grisms (probing 8800-15310\,\AA, rest-frame 4800--8500\,\AA\ for sources at $z\sim0.8$), both yielding a resolution of $R\sim700$. Individual exposures were 200\,s. NB$_K$ targets were observed for up to 8\,ks\,pix$^{-1}$, while NB$_J$ targets only required up to 2.5\,ks\,pix$^{-1}$ for similar S/N. Three telluric stars were observed per night at the closest possible air masses and positions to the targets. Darks, flats and arc frames were obtained at the beginning of each night. Across the four nights of observations weather conditions remained good with only some cirrus on the first night. Seeing was stable between 0.6\arcsec and 0.9\arcsec on the first three nights of the run, with a rise to 1.1\arcsec on the final night. The majority of measurements were taken with seeing $<$1\arcsec. Out of the 23 targets, we confirm 16 H$\alpha$ emitters: 8 at $z=0.84$ (NB$_J$ sample) and 8 at $z=2.23$ (NB$_K$ sample).

\subsection{Data Reduction: SofI, LIRIS and NICS}

SofI data were reduced using the SofI ESO pipeline version 1.5.4 and {\sc esorex} version 3.9.0 recipes. Briefly, master flat fields and master arc frames were produced per night, and frames were flattened. Initial wavelength calibrations were produced by matching the master arc frames with catalogued Xenon and Neon lines. The co-addition recipes corrected for distortion, crosstalk and slit curvature. We then sky-subtracted according to the ABBAAB jitter sequence and average-combined individual reduced frames. While {\sc esorex} provides a reasonable wavelength calibration, we improved upon it by matching $\sim50$ unblended OH lines. We used a polynomial fit for all our data-sets, and determined the coefficients by performing a least-squares fit on OH lines over a wide range of pixels that were detected on the science frames \citep[e.g.][]{Osterbrock.96}. This is consistent with the calibration derived from the arcs, but much more homogeneously spread across the observed spectral range. Standard deviations of residuals to the fits were checked to be random and at the level of $\sim4-6$\,\AA, the same order as our pixel scale.

The reduction of NICS and LIRIS data followed the same procedures and steps as for SofI, but the data were reduced with a customised set of {\sc python} scripts. All science frames were divided by master flat frames taken on the same night as their observation. Using the offsets of the jittering sequence and the declination of the field, pixel offsets were calculated and the spectra were average-stacked. We applied a clipping of the lowest- and highest-value pixels within each stack in order to eliminate hot pixels, cosmic rays and other potential artefacts. Some examples of the final 2D spectra are shown in Figure \ref{2D_SPECTRA}.

%
%
%
%
\begin{figure*}
\begin{tabular}{cccc}
\includegraphics[width=17cm]{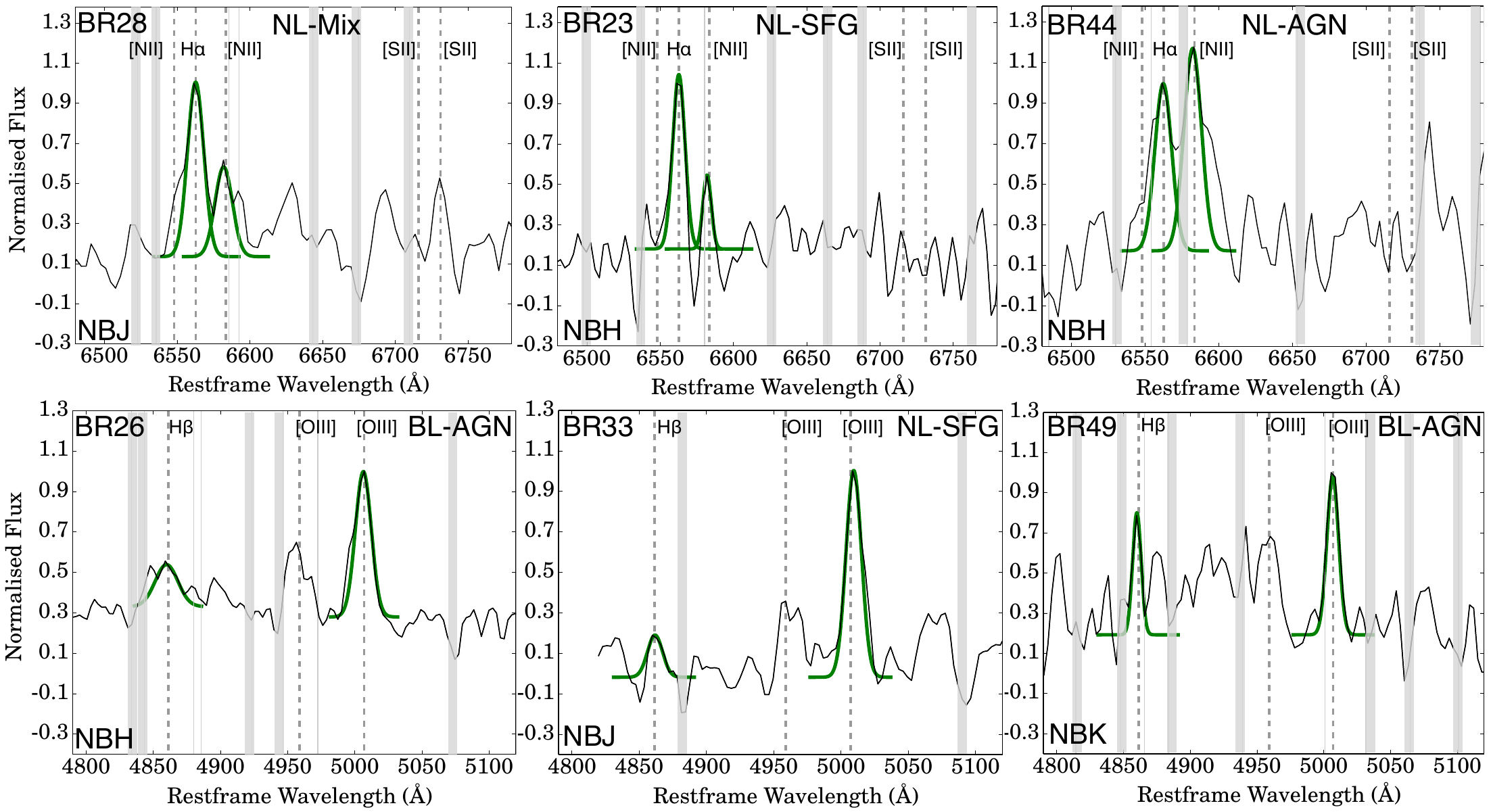}
\end{tabular} 
\caption{A selection of typical spectra showing, for three different sources (top), our coverage which allows us to trace H$\alpha$ and [N{\sc ii}] (and in some cases [S{\sc ii}]) and for another three sources (bottom) our coverage which allows us to trace H$\beta$ and [O{\sc iii}]. We find a variety of sources, but, in general, [O{\sc iii}] is almost always brighter than H$\beta$. Grey vertical lines indicate all OH lines (including weak OH lines) affecting our spectra green thick lines show our best Gaussian fits for H$\beta$, [O{\sc iii}], H$\alpha$ and [N{\sc ii}].}
\label{1D_SPECTRA}
\end{figure*}

%
%
%
%
\begin{figure*}
\begin{tabular}{ccc}
\centering
\includegraphics[scale=0.3]{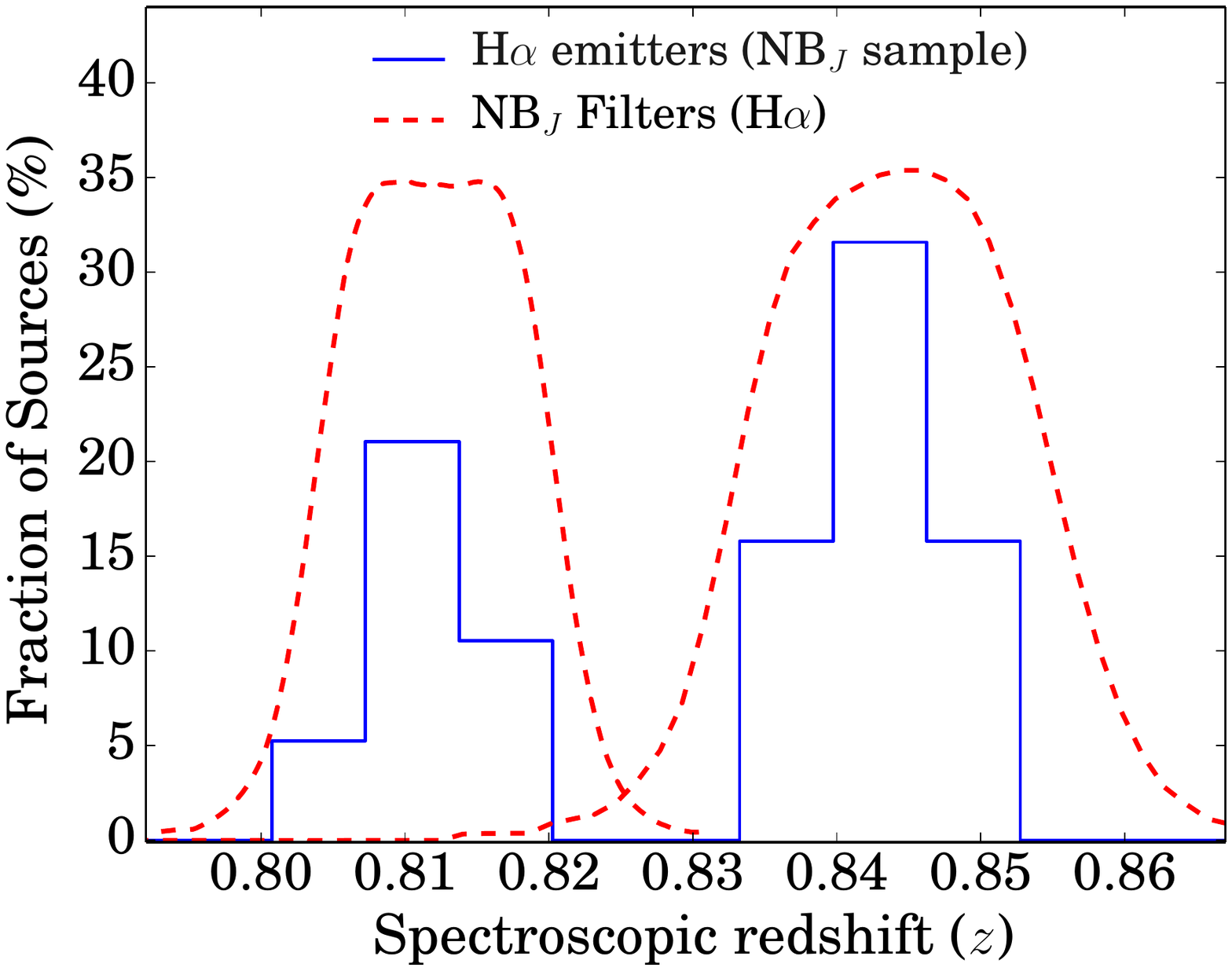}&
\includegraphics[scale=0.3]{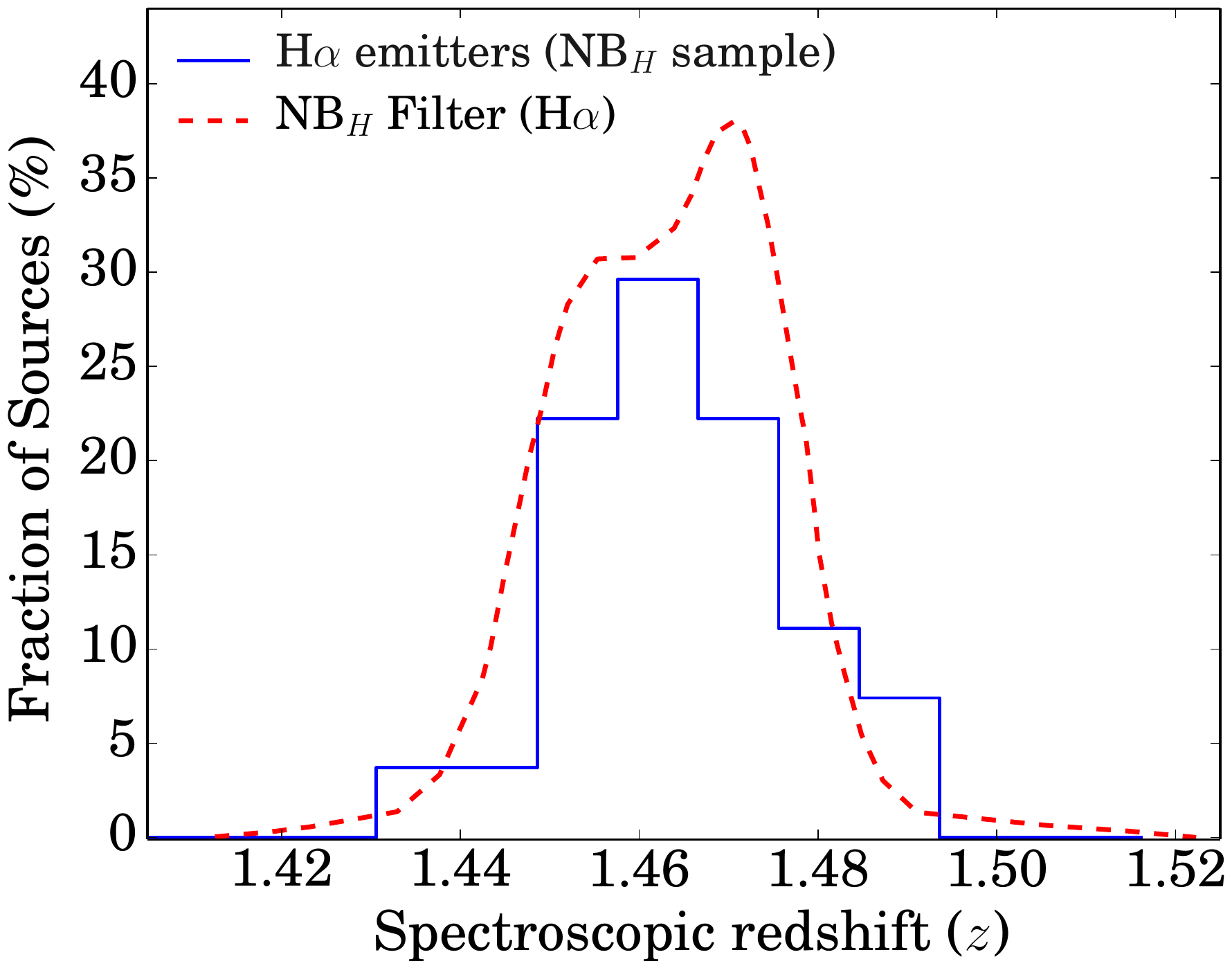}&
\includegraphics[scale=0.3]{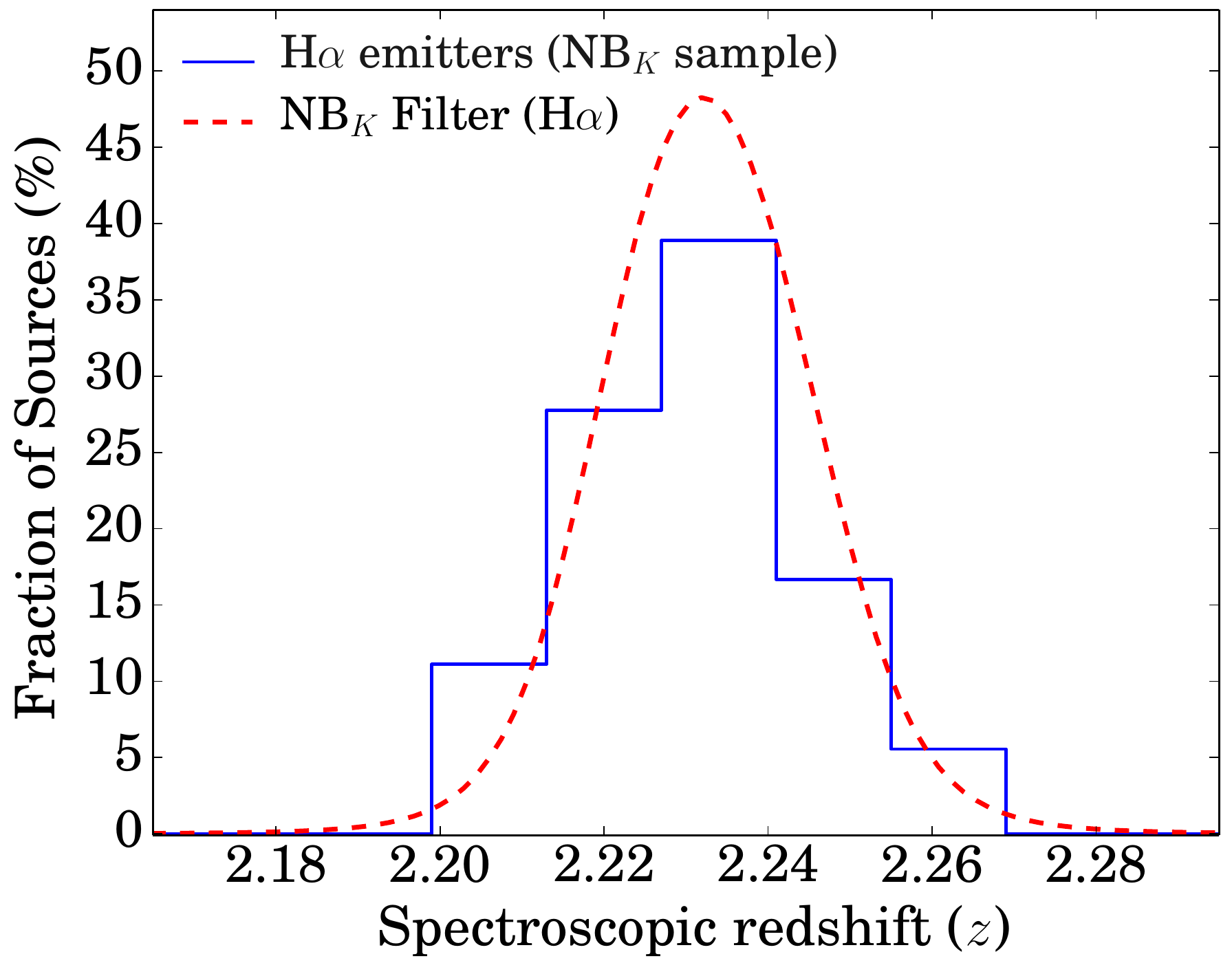}
\end{tabular}
\caption{\textit{Left-to-right}: The redshift distributions of of our NB$_{\rm J}$, NB$_{\rm H}$ and NB$_{\rm K}$ samples of H$\alpha$ emitters, respectively. The fraction of sources is simply the number of sources in each bin divided by the full sample at that redshift. In the case of NB$_{\rm J}$, the relative distributions between the two different narrow-band filters/data-sets is simply set by the number of followed-up sources in each data-set, as both had an equally high success rate. Over-plotted are the narrow-band profiles used for the selection of the samples. This shows that the redshift distribution of each sample follows the filter profile very well.}
\label{FILTER_PROFILES}
\end{figure*}

\subsection{Extraction and flux calibration}

For spectral extraction, whenever a small distortion across the detector was found, we first corrected for this gradient. We visually inspected each 2D spectrum (e.g. Figure \ref{2D_SPECTRA}) and extracted the 1D spectrum by summing up the pixels corresponding to $\sim1.5-2''$ in the spatial direction (we varied this slightly on a source by source basis to take into account the seeing variations and any important noisy features), corresponding to \til 15 kpc at all redshifts probed. Some typical examples are shown in Figure \ref{1D_SPECTRA}. Due to our strategy of acquiring a bright source and then rotating the instrument for the majority of the sources, we almost always have, together with our target, a bright source ($J\sim13-15$) typically 20-60\,$''$ away. These bright sources are also extracted in the same way, over the exact same aperture as our main science target (and any distortions corrected exactly in the same way and checked), and are flux-normalised by telluric spectra taken on the same night, in the same grism as the target spectrum and extracted over the same width.

\begin{figure}
\centering
\includegraphics[width=8cm]{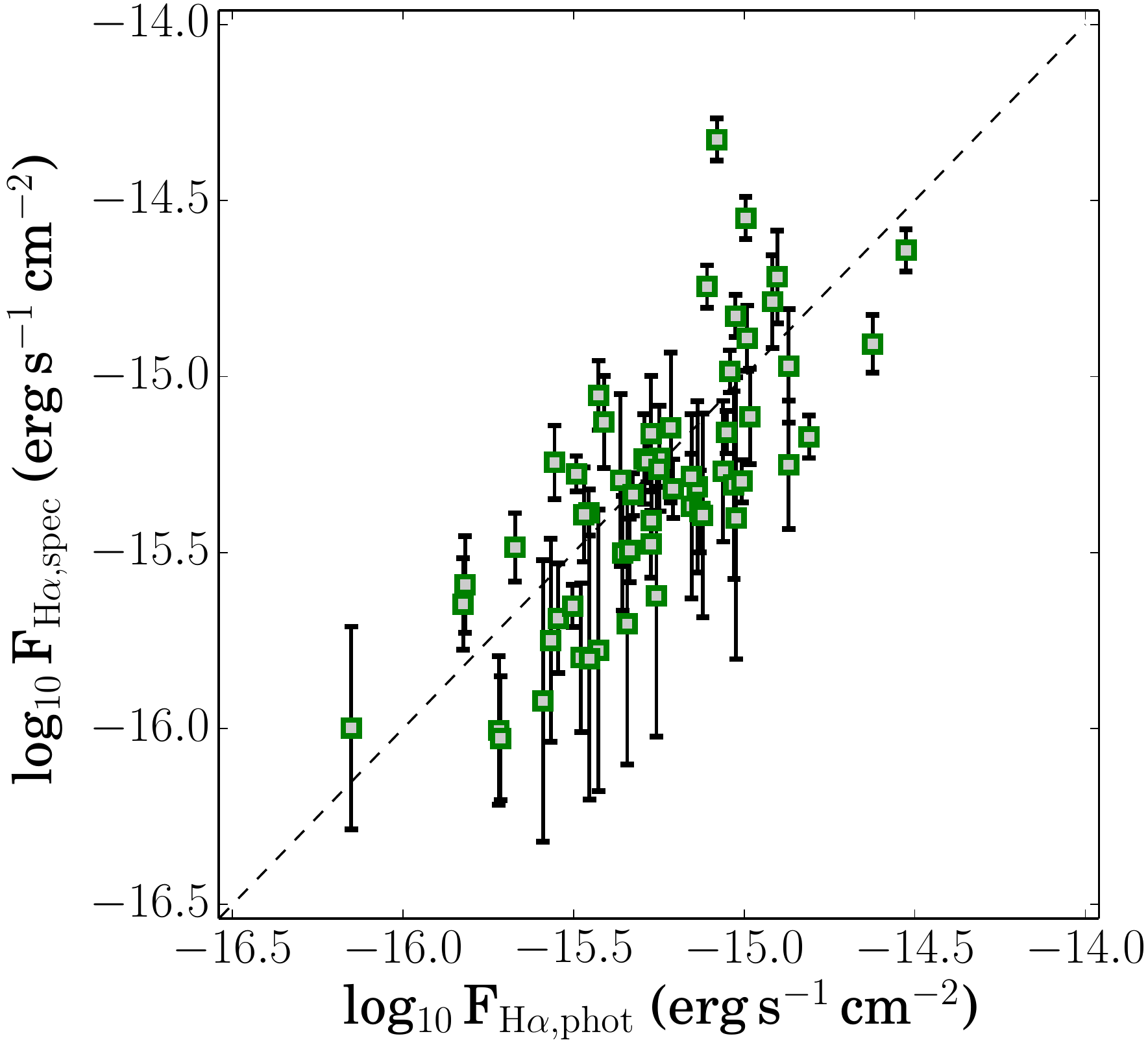}
\caption{A comparison between the spectroscopic H$\alpha$ fluxes and those derived from the narrow-band photometry. The dashed line shows the 1:1 relation. We find a good agreement between both, within the uncertainties, and without any strong biases/systematic offset.}
\label{fluxes_spec_phot}
\end{figure}

In order to estimate, and correct for, the light lost out of the slit, we use 2MASS photometry \citep{Skrutskie.06} and explore the wealth of relatively bright ($J\sim14$, thus yielding very high S/N for our exposure times) sources which we typically used to acquire our targets and that remained in the slit at all times. By using the known flux density of each of our bright sources ($J$ and $H$ or $H$ and $K$, depending upon grism used), we flux calibrate all our spectra. We note that this process assumes that the target and the bright source are equally well-centred in the slit, and of similar apparent angular extent: this is a good assumption for the sources we targeted.
We check that the flux calibration that we apply yields emission line fluxes that correlate well (and that have the same normalisation within the errors) with the estimates from the narrow-band photometry (see Figure \ref{fluxes_spec_phot}). Differences between NB estimated fluxes and spectroscopic fluxes are fully explained by either errors/uncertainties, redshifts (for some redshifts the filter profile has a lower transmittance, underestimating the flux, which can now be fully checked after determining the redshifts), and due to H$\alpha$ lines which are even broader than the narrow-band filter profile.

%
%
%
%

\section{Analysis} \label{section:MethMed}

\subsection{Line identification and spectroscopic redshifts} \label{sub:z}

We use both the 1D and 2D spectra in order to first identify the main emission line at the wavelength range covered by the narrow-band filter used to select each source. Out of our 73 targets, we identify a strong emission line in the vast majority of followed-up sources (64 of them, corresponding to a success rate of 88\%), with the remaining sources (9) being stars detected with very high S/N continuum and strong features in the near-infrared which mimic strong emission lines (although all these are easily classed as stars using colour-colour criteria, and thus none are in the \citealt{Sobral.13} samples). For the sources with an emission line, we produce redshift solutions, starting with identifying the emission line as H$\alpha$, but also assuming it can be any other strong emission line. We then look for further emission lines, exploring the wide wavelength coverage of all our spectra: we do this simultaneously in the 2D and 1D, and highlight the location of strong OH lines. Finally, after selecting the approximate correct redshift for each source, we fit Gaussian profiles to the main emission lines identified, and further refine the redshift and estimate the error on the redshift based on the standard deviation of redshifts obtained using each line individually. We find that out of the 64 emission line sources, 59 (92\%) are H$\alpha$ emitters, with the remaining being [O{\sc iii}] emitters and one low redshift emitter. As Figure \ref{FILTER_PROFILES} shows, the redshift distribution of H$\alpha$ emitters follows very closely what would be expected given the filter profiles and how efficient they should be at recovering H$\alpha$ (for broad H$\alpha$ the filter profiles are even sensitive to slightly higher and lower redshifts -- the filter profiles shown in the Figure assume a narrow H$\alpha$ line).

\begin{figure}
\includegraphics[scale=0.37]{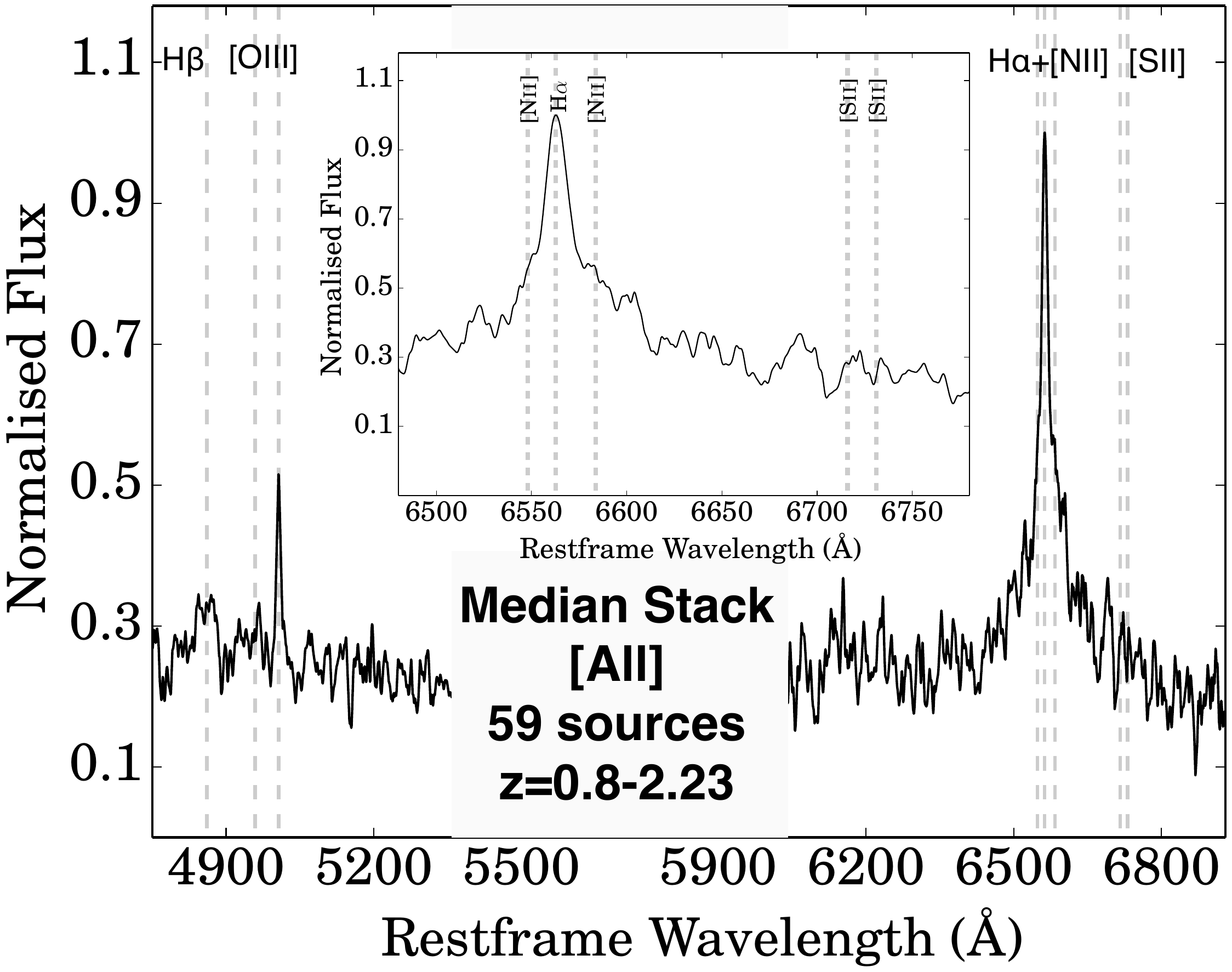}
\caption{The median stack of all 59 sources in our sample by normalising all sources by the peak of the H$\alpha$ emission line. This reveals a broad H$\alpha$, but that the resulting narrow-line profile dominates. The central region is masking the low S/N region which results from the very low atmospheric transmission between either $Y$ and $J$, $J$ and $H$ or $H$ and $K$ bands.}
\label{full_stack}
\end{figure}

There was no evidence of significant systematic offsets between the redshift determinations from our two strongest lines, H$\alpha$ and [O{\sc iii}]5007 (see e.g. Figure \ref{1D_SPECTRA}). For cases where we found only one line, within the boundaries of the narrow-band filter and not falling on a strong OH line, it was assumed to be H$\alpha$ (provided it was consistent with the lack of other lines). We check that all these single-line sources have photometric redshifts and colours consistent with being H$\alpha$ emitters \citep[e.g.][]{Sobral.13}. Table \ref{RESULTS} presents the full details on the number of sources and the main emission lines detected which will be used to classify the sources. By normalising at the peak of the H$\alpha$ emission line, we also median stack all the sources. Figure \ref{full_stack} shows the results.

%
%
%

\subsection{Line measurements and samples} \label{sub:line_meas}

\subsubsection{Main emission lines}

Our observations covered the wavelength range $\sim0.9--2.52\,\mu$m in order to probe the rest-frame optical. Our main lines of interest are H$\beta$ (4861\AA), the [O{\sc iii}] doublet (4959\AA, 5007\AA), the [N{\sc ii}] doublet (6548\AA, 6584\AA) and H$\alpha$ (6562.8\AA). For the remaining of the paper, we refer to [O{\sc iii}]\,5007\AA~ and [N{\sc ii}]\,6584\AA~  as [O{\sc iii}] and [N{\sc ii}] respectively. By using the redshift of each source and its error, and the location of each strong OH line, we fit Gaussian profiles to each emission line, after removing the continuum with two linear relations which are calculated independently at the red and at the blue sides of each emission line, by also excluding any nearby emission lines and/or strong OH lines. Whenever we fail to detect an emission line with $>2$\,$\sigma$, we assign it an upper limit of 2$\sigma$. For H$\alpha$ we fit simultaneously a narrow (typically a few 100\,km\,s$^{-1}$, comparable to the spectral resolution, $\sim100-200$\,km\,s$^{-1}$) and a broad (typically a few 1000\,km\,s$^{-1}$) Gaussian profile, in an automated way, and without applying any correction for the spectral resolution, as we are mostly interested in distinguishing between broad and narrow lines within the same data-set. We also measure line profiles manually, source by source, and check that the results are fully consistent within the errors. Other detected lines in our spectra included H$\gamma$, He{\sc ii} and the [S{\sc ii}] doublet, but only in broad-line AGN, and these lines are not used in the analysis. Gaussian fits of the emission lines were integrated to obtain line fluxes.

\subsubsection{Low S/N sample}

For a fraction of our sources (24 sources; 41\,\%), only one single narrow-line is detected, which we assume is H$\alpha$. The typical H$\alpha$ S/N for these 24 sources is $\sim2.5-4.5$. These sources are found at the lowest fluxes, with an average flux ($4\times10^{-16}$\,erg\,s$^{-1}$\,cm$^{-2}$) which is $\sim2$ times lower than the high S/N sample (\S\ref{highSN}). It is not possible to further investigate the nature of these apparent narrow-line emitters individually. However, in order to further constrain their nature as a population, we stack the spectra of all these 24 sources. We do not detect [N{\sc ii}], implying a low [N{\sc ii}]/H$\alpha<0.15$, consistent with photo-ionisation by star formation \citep[e.g.][]{Baldwin.81, Rola.97, Kewley.13}, and we find [O{\sc iii}]/H$\beta$ $\sim5$. This probably implies that the majority of the unclassified galaxies are metal-poor star-forming galaxies. Thus, while we cannot constrain the nature of these sources individually, we keep these sources for the remaining of the analysis, assuming that the bulk of them are not AGN, in agreement with e.g. \cite{Stott13b} at even lower fluxes, and also with what we find in \S\ref{lowLUM_HA}.

\subsubsection{High S/N sample}\label{highSN}

As we are particularly interested in unveiling the nature of the most luminous H$\alpha$ emitters, out of the full sample for which we confirmed and obtained a spectroscopic redshift, we apply a S/N$\,>5$ cut on the H$\alpha$ emission line. This allows us to obtain a  sub-sample of 35 luminous H$\alpha$ emitters for which we can further constrain their nature. Table \ref{RESULTS} provides information on the full sample and on how many sources have information available for the different lines.

%
%
%
\begin{table*}
\caption{Number of sources in our sample. We first present the full number of sources with spectroscopic redshifts, then the sources with high enough signal to noise to obtain more information. The number of broad- and narrow-line H$\alpha$ sources are only provided for the high S/N sources, where one can clearly distinguish between both -- all sources with S/N$<5$ have a narrow H$\alpha$ emission line, but the S/N is simply not sufficient to see any potential broad component. We then present the number of sources for which we are able to determine line ratios (we label as ``BPT 4 lines" the sources for which we can determine both [N{\sc ii}]/H$\alpha$ and [O{\sc iii}]/H$\beta$), and those we classify as AGN and SFG. For the unclassified sources, we also show, in brackets, the number of sources which have high S/N at H$\alpha$, but for which it is not possible to classify them, either because they have line ratios that place them between SFGs and AGNs, or, in the case of 6 sources at $z=0.8$, because of the lack of blue coverage -- the [N{\sc ii}]/H$\alpha$ ratios of those sources also do not allow to clearly classify any of them as AGN. Unclassified sources are likely to be star-forming dominated.}
\label{RESULTS}
\begin{tabular}{ccccccccccc}
\hline
Sample  & z$_{\rm spec}$  & S/N$<5$ & S/N$>5$  & BL H$\alpha$ & NL H$\alpha$ & NL [NII]/H$\alpha$ & BPT 4lines  & SFG  & AGN & Unclassified \\
\hline
$z=0.8$ & 17 & 6 &  11 & 1  & 10 & 9 & 4 & 3 & 1 & 13 (7)  \\ 
$z=1.5$ & 28 & 9 &  19 & 10  & 9 & 9 & 8 & 3 & 14 & 11 (2)  \\
$z=2.2$ & 14 & 9  & 5 & 3  & 2 &  2 & 2 & 1 & 3 & 10 (1)  \\
\hline
All & 59 & 24 & 35 & 14  & 21 & 20 & 14 & 7  & 18 & 34 (10)  \\
Fractions & 100\% & 41\% & 59\% & 24\% & 36\%  & 34\% & 24\% & 12\% & 30\% & 58\% (17\%)  \\
\hline 
\end{tabular}
\end{table*}

\subsubsection{H$\alpha$ FWHM: identifying broad-line AGN} \label{subsub:blnl}

%
%
%
\begin{figure}
\includegraphics[width=8cm]{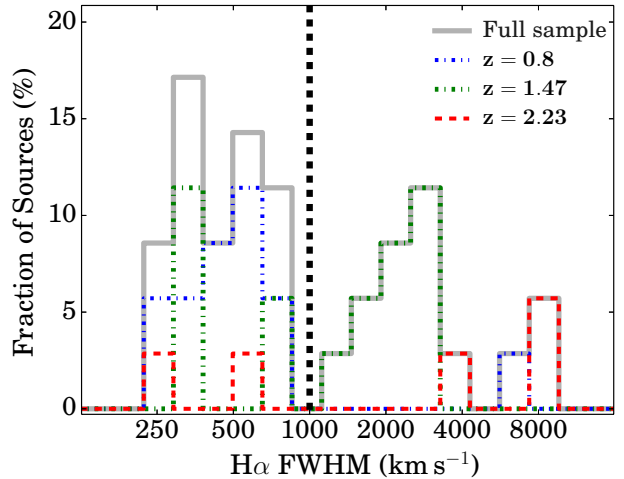}
\caption{The distribution of H$\alpha$ FWHMs for our high S/N sample (35 sources, see Table \ref{RESULTS}) in the three redshift ranges we probed (vertical dashed line shows the separation we adopt to differentiate between narrow- and broad-line H$\alpha$ emitters). We find that the typical narrow-line H$\alpha$ emitters have FWHM of $\sim300-400$\,km\,s$^{-1}$ and that these dominate the sample overall, although they are the faintest emitters within our sample -- at higher luminosities higher FWHM dominate. The broadest H$\alpha$ lines are found at $z=2.23$ (see also Figure \ref{2D_SPECTRA}). The fraction of broad line emitters is the highest at $z=1.47$.}
\label{fig:FWHMs}
\end{figure}

Very broad H$\alpha$ emission with high FWHM (typically $>1000$\,km$^{-1}$) can be seen as a clear and robust indication of AGN activity: broad-line AGN (BL-AGN). Here we use a rest-frame H$\alpha$ FWHM of $>1000$\,km\,s$^{-1}$ to distinguish between what we will henceforth refer to as broad- and narrow-line emitters, which is consistent with the relevant literature \citep[e.g.][]{Stirpe.90,Ho.97}. Broad-line emitters are hereafter assumed to be AGN, since there are few processes other than gravitational motions close to a central black hole that can account for such broadening in a galactic spectrum. For example, strong outflows in massive star-forming galaxies at $z\sim2$ lead to FWHMs of $\sim450$\,km\,s$^{-1}$ \citep[][]{Newman2012}. Much broader emission lines, in excess of 1000\,km\,s$^{-1}$ are seen in central parts of massive galaxies at $z\sim2$, attributed to AGN activity \citep[][]{Genzel2014}. Starburst-driven galactic winds may be able to drive gas to velocities up to $\sim3000$\,km\,s$^{-1}$ \citep{Heckman.03}, but this would result in highly asymmetric emission line profiles. Although we find tentative evidence for some asymmetry in some of the broader lines (blue-shifted), this seems to be on top of a broad, symmetric, BL-AGN H$\alpha$ profile.

We find 14 broad line AGN out of our sample of 59 H$\alpha$ emitters (24\% of the full sample), 1 at $z\sim0.8$, 10 at $z\sim1.47$ and 3 at $z=2.23$. This already reveals that there is a significant fraction of BL-AGN at the highest H$\alpha$ luminosities at $z\sim0.8-2.23$ and a higher broad-line AGN fraction at $z=1.47$. Among our BL-AGNs, two stand out in particular, as their H$\alpha$ FWHM $>$ 10$^4$ kms$^{-1}$, or about 0.03$c$ (see Figure \ref{fig:FWHMs} for the full distribution of FWHMs). These are BR-60 and BR-64, both at $z\sim2.2$, shown in Figure~\ref{2D_SPECTRA}.

In Figure \ref{fig:FWHMs} we show the distribution of H$\alpha$ FWHMs for our high S/N sample and also for sub-samples at each redshift. Narrow-line H$\alpha$ emitters (H$\alpha$ FWHM $\le$ 1000\,kms$^{-1}$) dominate the $z\sim0.8$ distribution, but are still significant contributors to the $z\sim1.5$ and $z\sim2.2$ distributions. We note that lower S/N sources not shown in Figure \ref{fig:FWHMs} are consistent with being narrow-line emitters (the stack reveals a narrow H$\alpha$ line $\sim400$\,km\,s$^{-1}$). We further note that we may miss weak BL components, particularly in the lower S/N spectra, and thus BL fractions should conservatively be interpreted as lower limits.

%
%
%
\begin{figure}
\centering
\includegraphics[width=8cm]{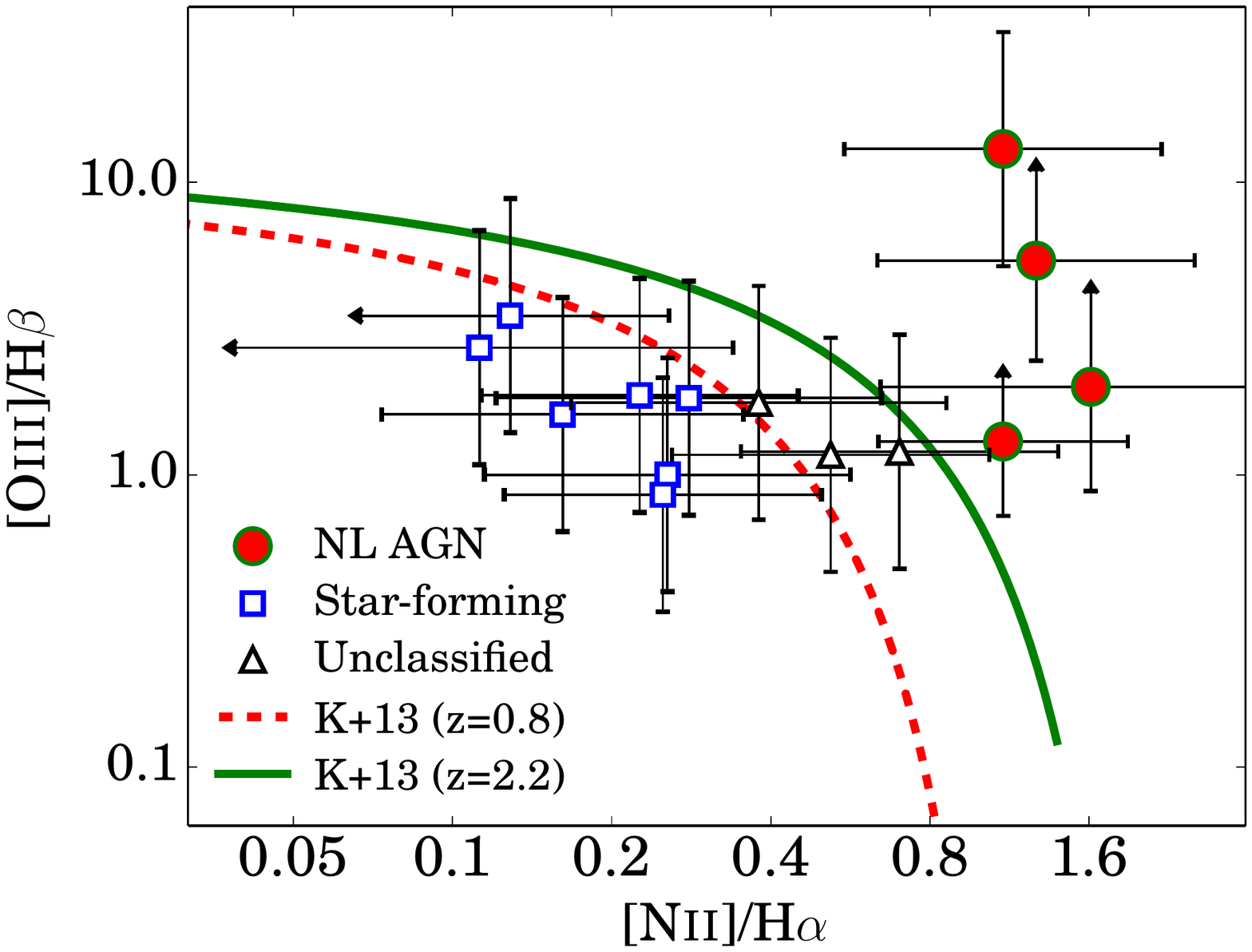}
\caption{SFGs and narrow-line AGNs can be distinguished from one another, for targets that exhibit narrow-line H$\alpha$ emission, by the line ratios [N{\sc ii}]/H$\alpha$ and [O{\sc iii}]/\hb \citep{Kewley.13}. We show the line ratios of targets from our NB$_{\rm J}$, NB$_{\rm H}$ and NB$_{\rm K}$ samples. The boundaries between these two populations are shown for the lowest and highest redshifts in the sample from \citet{Kewley.13}, and the classification between AGN or star-forming takes into account the redshift. Error bars are the 1$\sigma$ uncertainties in the measurements. Since the purpose here is to distinguish between NL AGN and NL SF, and because it is not possible to reliably estimate the narrow-line [N{\sc ii}]/H$\alpha$ ratio for the broad-line AGN (due to resolution), we do not show the 14 BL AGN in our sample.}
\label{fig:SFG_vs_AGN}
\end{figure}

%
%
%
\begin{figure*}
\centering
\begin{tabular}{ccc}
\includegraphics[width=8cm]{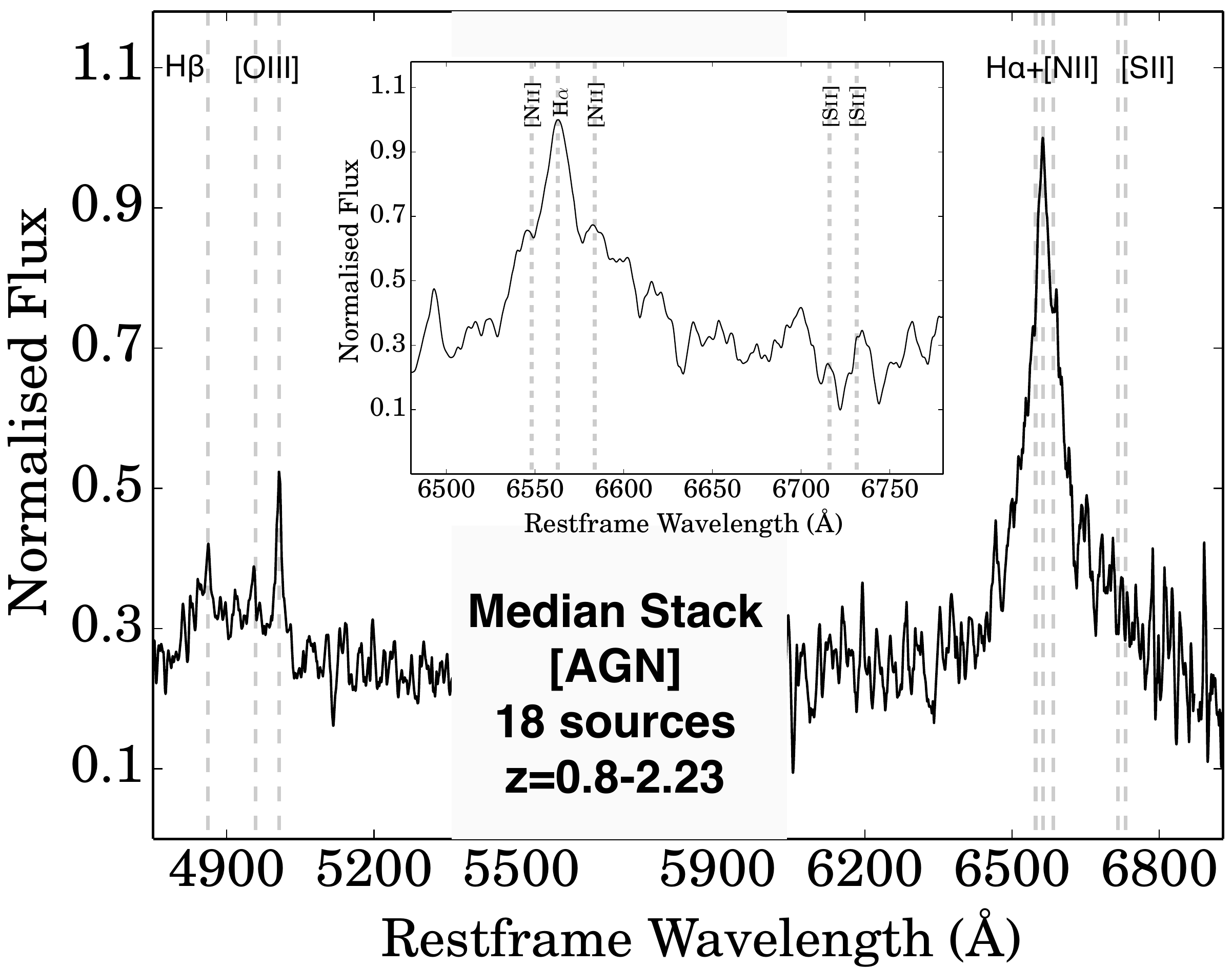}&
\includegraphics[width=8cm]{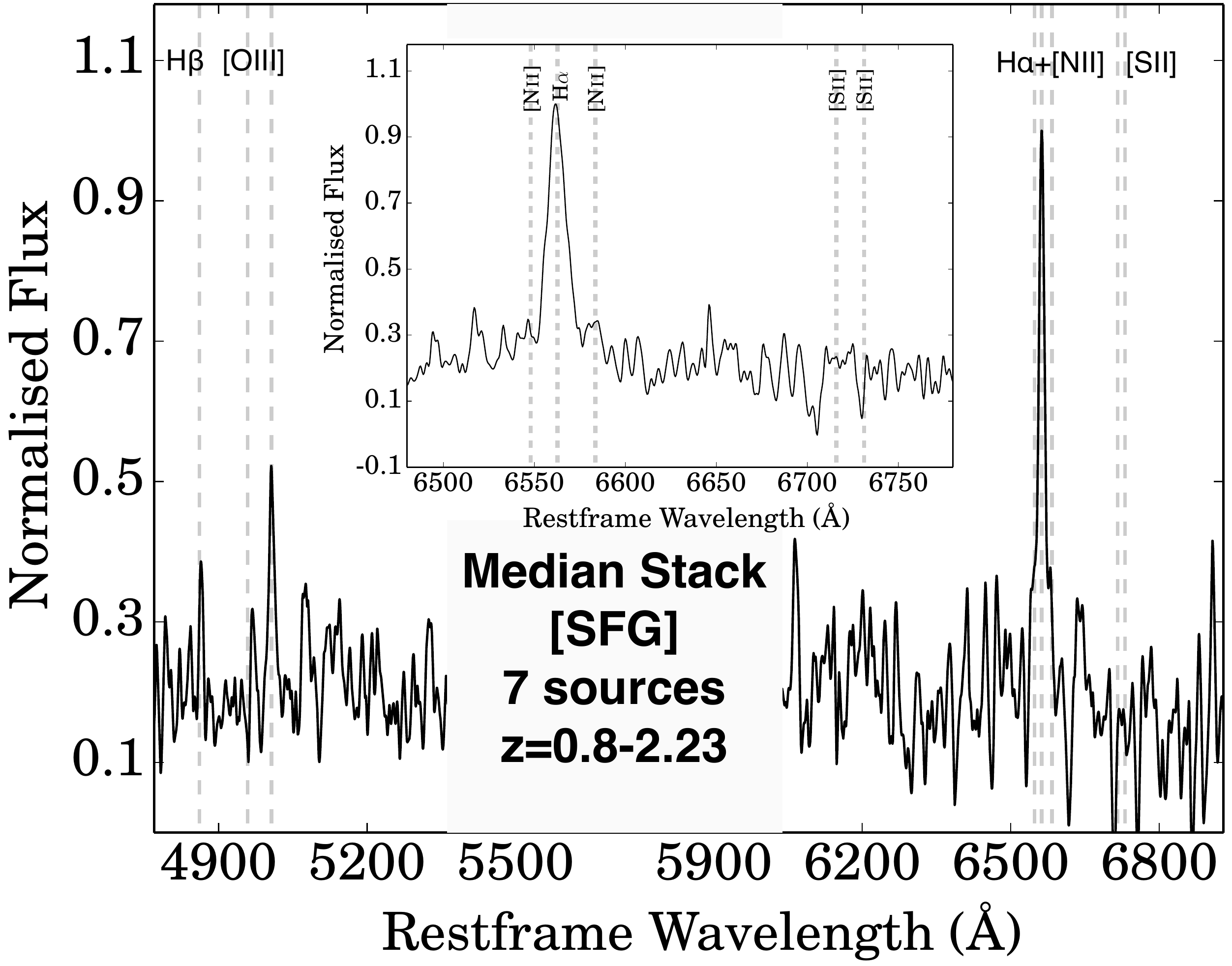}\\
\end{tabular}
\caption{$Left$: The median stacked spectrum of all AGN in our sample (18 sources used in the stack). The stacks reveals a strong H$\alpha$ line which can only be described fully with a combination of 2 Gaussian profiles: a dominating profile of $\sim4000$\,km\,s$^{-1}$, and an even broader profile of $\sim10000$\,km\,s$^{-1}$ which also shows some asymmetry suggesting blue-shifted emission. These reveal a range in black hole masses within our sample, but show that these AGN are typically very massive, and likely able to drive powerful outflows. We also find broad H$\beta$ in the stack. $Right$: The median stack for star-forming galaxies (7 sources used in the stack), showing strong H$\alpha$, very weak [N{\sc ii}] detections and detections of [O{\sc iii}] and H$\beta$ fully consistent with typical star-forming galaxies. The H$\alpha$ emission line is well fitted with a Gaussian profile and FWHM$\sim300$\,km\,s$^{-1}$. We find a [N{\sc ii}]/H$\alpha$ ratio of 0.05$\pm$0.02, implying that the median metallicity of our luminous star-forming H$\alpha$ emitters at $z\sim0.8-2.2$ is 12\,+\,log(O/H)\,=8.16$\pm$0.08. Our limit on the [S{\sc ii}]/H$\alpha$ line ratio also implies a very high ionisation potential, again consistent with very low metallicity and very high luminosities. Note that the relatively low S/N H$\beta$ detection is also driven by the H$\beta$ line being strong affected by strong OH lines for the bulk of the sample.}
\label{fig:stacked_subsamples}
\end{figure*}

\subsection{Distinguishing between NL AGN and SFGs} \label{sub:AGNvsSFdiag}

Out of the full sample of 59 H$\alpha$ emitters, we assume our low S/N sample (24 sources) are SFGs. For the remaining 35 sources, we already found that 14 are BL-AGN. We now attempt to classify the remaining 21 high S/N sources, which are all narrow-line emitters, as star-forming galaxies or narrow-line AGN (NL-AGN). This can be done using emission line ratios \citep[e.g.][]{Baldwin.81, Rola.97, Kewley.13}. However, the separation between AGN and typical star-forming galaxies has been shown to evolve with redshift \citep[see e.g.][]{Shapley2015}, and thus we use the \cite{Kewley.13} parameterisation -- although we note that such work is currently mostly theoretical, while observations are starting to provide very useful constraints. Figure~\ref{fig:SFG_vs_AGN} illustrates the use of the \cite{Kewley.13} diagnostic for distinguishing the nature of narrow-line emitters. If we do not detect \hb at more than 2$\sigma$ significance due to being affected by a strong OH line, we use the measured limit (3 sources, all AGN), but show those as lower limits. In two (2) cases [N{\sc ii}] is below 2\,$\sigma$. For those we assign the 2$\sigma$ limit as the [N{\sc ii}] flux (but we also plot those as upper limits), and those are the sources with the lowest [N{\sc ii}]/H$\alpha$ in our sample ($\sim0.1$) and are clearly star-forming. Table \ref{RESULTS} provides the full information regarding the availability of each of the line ratios, the samples, and the results in the classification of sources. We also median stack all sources, after normalising them to peak H$\alpha$ emission, that we classify as AGN and all the sources we classify as SFGs using the BPT: we show the stacks in Figure \ref{fig:stacked_subsamples}.

\subsection{Lower Luminosity H$\alpha$ emitters} \label{lowLUM_HA}

In order to estimate the AGN fraction among lower luminosity/more typical H$\alpha$ emitters, and compare with our luminous H$\alpha$ emitters, we explore the general HiZELS sample \citep[][]{Sobral.13}, which allows us to probe the same redshift ranges as in this study, with the same selection. We use the results from \cite{Stott13b} that followed-up a sample of typical H$\alpha$ emitters from \cite{Sobral.13} with FMOS/Subaru, finding an AGN fraction of about $\sim11$\%. Within the uncertainties, more typical H$\alpha$ emitters (with lower luminosities) have a much lower AGN fraction than those studied in this paper. This is in good agreement with \cite{Garn.10}.

We also use the results from \cite{Calhau15} for more details on AGN activity for more typical H$\alpha$ emitters within the HiZELS data-set. Briefly, deep Spitzer/IRAC data is used to search for red colours beyond $\sim$ 1.6\,$\rm \mu$m rest frame. A clearly red colour indicates the presence of hot dust and of an AGN, while typical star-forming galaxies reveal a blue colour beyond 1.6\,$\rm \mu$m rest frame. 

For $z=0.8$, we use [3.6]-[4.5] in order to identify AGN, while for $z=1.47$ we use [4.5]-[5.8] and use [5.8]-[8.0] for $z=2.23$. Specifically, we use the colour selections [3.6]-[4.5]$>$0.0 for $z=0.8$, [4.5]-[5.8]$>$0.15 for $z=1.47$ and [5.8]-[8.0]$>$0.3 for $z=2.23$. These cuts take into account the distribution of sources and the increase in the scatter of the colour distributions, but are also motivated to select {\it Chandra} and VLA detections, indicative of AGN activity. This results in a $10\pm5$\% AGN fraction at $z=0.8$, $16\pm5$\% at $z=1.47$ and $15\pm4$\% at $z=2.23$ consistent with little to no evolution, particularly as the samples at higher redshift probe higher H$\alpha$ luminosities. 

Overall, the results clearly show that at $z\sim0.8-2.23$, the AGN fraction of low luminosity H$\alpha$ emitters ($\le L_{\rm H\alpha}^*$) is at a level of $\sim10-15$\% (and certainly below 20\,\%), much lower than that of much higher luminosity H$\alpha$ emitters. We also do not find any significant evidence for redshift evolution.

\section{Results and Discussion} \label{sec:Res}

For our full sample of 59 H$\alpha$ emitters, we have 24 low S/N sources, which we are unable to classify, but that are likely SF dominated. For the remaining 35 sources (the high S/N sample), we find 14 BL-AGN, 4 NL-AGN (thus, 18 AGN), 7 star-forming galaxies and 3 sources which are unclassified. We thus find an AGN fraction of $\sim30$\% among the full sample of 59 H$\alpha$ emitters (see Table \ref{RESULTS}), and a $\sim50$\% AGN fraction among the high S/N sample.

%
%
%
\begin{figure*}
\includegraphics[width=17.5cm]{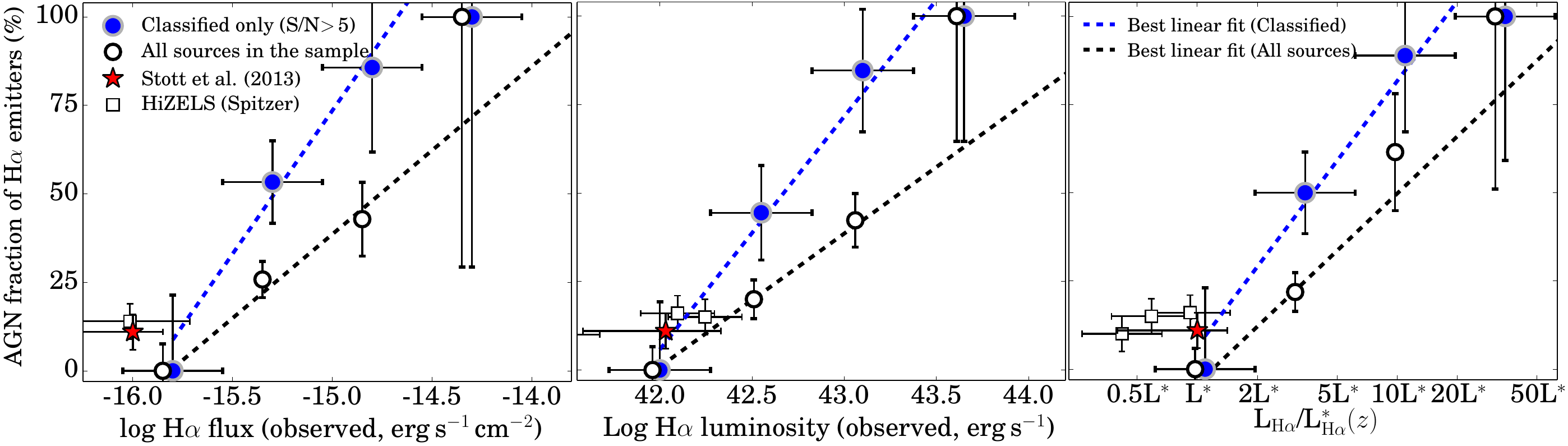}
\caption{$Left$: The fraction of AGN as a function of H$\alpha$ observed flux when considering only directly classified sources and for the full sample (including the lower S/N sources; bins shifted by $-$0.05 so they do not overlap). We also show the best linear fits (see Table \ref{best_fits}). This shows a strong increase in the fraction of AGN for higher fluxes, dominated by the rise of the fraction of broad-line AGN. However, high luminosities at high redshift will be observed as lower observed fluxes. $Middle$: The fraction of AGN as a function of H$\alpha$ luminosity. We find that the AGN fraction is higher at the highest H$\alpha$ luminosities. However, as shown in \citealt{Sobral.13}, the typical luminosity of H$\alpha$ emitters evolves significantly with cosmic time. $Right$: By taking into account the evolution of $L^{*}$, we recover a strong relation between the location within the H$\alpha$ luminosity at each redshift, and the AGN fraction. This shows that while around $L^{*}$ and lower H$\alpha$ luminosities only a minor fraction of H$\alpha$ selected sources are AGN, the fraction rises steeply for higher luminosities. The \citet{Stott13b} study observed and characterised lower-luminosity HiZELS sources shown here for comparison, as it extends our results to lower fluxes and lower values. For comparison, we show the AGN fraction estimated for the HiZELS sample, as detailed in \S\ref{lowLUM_HA}. We also show the best linear fits and Table presents the best fit coefficients and errors (1$\sigma$).}
\label{fig:lha_lhastar} 
\end{figure*}

\subsection{Broad Line H$\alpha$ emitters: number densities and black hole masses}

Using the measured H$\alpha$ FWHMs, H$\alpha$ luminosities and Eq. 9 from \cite{Greene.05}, we may obtain an estimate of the black-hole (BH) masses of the AGNs in our sample. The average BH mass across all AGNs in our survey is $\sim10^{8\pm1}$M$_{\odot}$, with a relatively high standard deviation mainly coming from larger-than-average masses of the broadest BL-AGNs in the NB$_{\rm K}$ ($\sim10^9$M$_{\odot}$; see Figure \ref{2D_SPECTRA}) sample. We note that the estimation of black hole masses from line widths is only valid for cases where we can see the broad line region, and thus we restrict our analysis to those. This is because the estimate is based on simple circular motion arguments, thus the need to estimate velocity and radius. We compare our measurements with \cite{HeckBest2014}, to find that many of our BL-AGN are relatively ``normal" AGN \citep[][]{HeckBest2014}, with masses of a few times $10^7$\,M$_{\odot}$, although two of our BL-AGN reach masses more typical of quasars at $z\sim2$ \citep[e.g.][]{McLure04} with $10^9$\,M$_{\odot}$.

Over all redshifts, we find that the volume density of BL-AGN among luminous H$\alpha$ emitters (for volumes where we are spectroscopically complete, thus we do not apply any correction for incompleteness) is 5.7$\pm1.5\times10^{-6}$\,Mpc$^{-3}$ (3$\pm3\times10^{-6}$\,Mpc$^{-3}$, 9$\pm3\times10^{-6}$\,Mpc$^{-3}$ and 3$\pm2\times10^{-6}$\,Mpc$^{-3}$ at $z=0.8$, $z=1.47$ and $z=2.23$, respectively). Our results are therefore consistent with a constant volume density of broad line AGN at the peak of AGN and star-formation activity, of roughly $\sim6\times10^{-6}$\,Mpc$^{-3}$, but with a potential peak at $z\sim1.5$. These number densities are roughly consistent with the number density of massive BL-AGN \citep[e.g.][]{McLure04}, given the estimates of black hole masses for our BL-AGN: $\sim10^{-6}-10^{-5}$\,Mpc$^{-3}$. As mentioned in \S3.2.4, we note that we may miss weak BL components, particularly in the lower S/N spectra, and thus our number density of massive BL-AGN should conservatively be interpreted as lower limits. While our sample of BL-AGN is too small to further split it per redshift, our findings are consistent with a decrease in the BL-AGN fraction for fixed H$\alpha$ luminosity, with increasing redshift.

\subsection{Evolution of AGN fraction with H$\alpha$ flux, luminosity, cosmic-normalised luminosity and redshift} \label{subsub:AGN_LHA}

Here we investigate how the fraction of AGN among H$\alpha$ emitters varies with H$\alpha$ flux, luminosity and $L_{\rm H\alpha}$/$L_{\rm H\alpha}^*(z)$, for our full sample, and when we restrict the sample to only sources we can individually classify. We also provide the best linear fit for each of the relations we find (see Table \ref{best_fits}).

As the left panel of Figure \ref{fig:lha_lhastar} shows, the AGN fraction rises significantly with increasing observed H$\alpha$ flux. This is seen both when we use the full sample, and the sample of classified sources only. This is mostly driven by the bright BL-AGN which, even at higher redshift ($z\sim2.2, 1.47$), are able to produce observable fluxes which are still much higher than more typical star-forming galaxies at $z\sim0.8$. 

We also find a strong correlation between the AGN fraction and H$\alpha$ luminosity, shown in the middle panel of Figure \ref{fig:lha_lhastar}. However, given that the typical H$\alpha$ luminosity is strongly increasing with look-back time/redshift, we also look for a potential correlation between the AGN fraction and the cosmic-normalised H$\alpha$ luminosity, which is simply $L_{\rm H\alpha}$ at a redshift $z$ divided by $L_{\rm H\alpha}^*(z)$ by using the results presented in \cite{Sobral.13}. Similar uses of this normalised quantity can be seen in e.g. \cite{Sobral.10} and \cite{Stott13a}. As the right panel of Figure \ref{fig:lha_lhastar} clearly shows, there is a strong correlation between AGN fraction and how luminous an H$\alpha$ emitter is relative to the typical H$\alpha$ luminosity ($L_{\rm H\alpha}^*$) at its cosmic time. The AGN fraction measured by \cite{Stott13b}, and those by \cite{Garn.10}, of much more typical H$\alpha$ emitters from the same survey, also fully agree with this trend. Our further investigation also shows that at $L^*$ and below, at all the redshifts probed, the AGN fraction is $\sim10-15$\%. However, as our results show, the AGN fraction rises with increasing $L/L_{\rm H\alpha}^*$, becoming $\sim25$\% by $\sim2L_{\rm H\alpha}^*$, 50\% by $\sim5L_{\rm H\alpha}^*$ and becoming essentially 100\% by $\sim50L_{\rm H\alpha}^*$, the most luminous sources in our survey.

%
%
\begin{table}
\caption{Best-fit linear relation as a function of different quantities/properties (all in $\log_{10}$ form): H$\alpha$ observed flux, H$\alpha$ observed luminosity and L$_{\rm H\alpha}$/L$_{\rm H\alpha}^{*}(z)$. We provide parameters $A$ and $B$ for each property/quantity, $x$ (AGN fraction= $Ax+B$), including the 1\,$\sigma$ error for each parameter when fitting both simultaneously.}
\begin{tabular}{ccc}
\hline
Property - sample	&	A			&	B 	 \\	
\hline						
H$\alpha$ flux (log$_{10}$) - All	 & $0.47\pm0.13$ & $7.5\pm2.0$   \\
H$\alpha$ flux (log$_{10}$) - S/N\,$>5$ 	&  $0.81\pm0.27$ & $12.9\pm4.1$  \\
\hline
H$\alpha$ luminosity (log$_{10}$) - All	 & $0.38\pm0.09$ & $-15.8\pm3.6$   \\
H$\alpha$ luminosity (log$_{10}$) - S/N\,$>5$	&  $0.66\pm0.19$ & $-27.6\pm8.2$  \\
\hline
L$_{\rm H\alpha}$/L$_{\rm H\alpha}^{*}(z)$ (log$_{10}$) - All & $0.54\pm0.13$ & $-0.04\pm0.06$ \\
L$_{\rm H\alpha}$/L$_{\rm H\alpha}^{*}(z)$ (log$_{10}$) - S/N\,$>5$ & $0.75\pm0.19$ & $0.06\pm0.10$ \\
\hline
\end{tabular}
\label{best_fits}
\end{table}

We test the statistical significance of the trends that we observe, particularly to evaluate which is the best predictor of the AGN fraction: H$\alpha$ flux, luminosity, or $L/L_{\rm H\alpha}^*(z)$. We use our binned data to find that all trends (with flux, luminosity and $L/L_{\rm H\alpha}^*(z)$) are significant at $>3$\,$\sigma$ on their own (comparing to no relation, i.e., a constant), considering only the classified sources (and considering all sources in brackets); $3.3(5.5)\sigma$, $3.1(5.1)\sigma$ and $4.6(6.4)\sigma$, respectively for H$\alpha$ flux, luminosity and $L/L_{\rm H\alpha}^*$ -- revealing that the AGN fraction correlates most strongly with $L/L_{\rm H\alpha}^*(z)$, for both the classified sources and when using the entire sample. Including the data-point from \cite{Stott13b} increases the significance of the trends by about $1\sigma$, but differences are maintained.  We further investigate the significance of the trends we find by binning the data 100,000 times with a range of random bin centres and bin widths within the parameter space that we probe. The results confirm that there is a significant relation between the AGN fraction and H$\alpha$ flux, luminosity and $L/L_{\rm H\alpha}^*(z)$, with all fits being at least $5\sigma$ away from no relation. We also find that the correlation is always more significant with $L/L_{\rm H\alpha}^*(z)$. We therefore conclude that while the three quantities are good predictors of the AGN fraction, for our probed parameter range, $L/L_{\rm H\alpha}^*(z)$ is the best.

%
%
%
%
%
\begin{figure*}
\includegraphics[width=15.5cm]{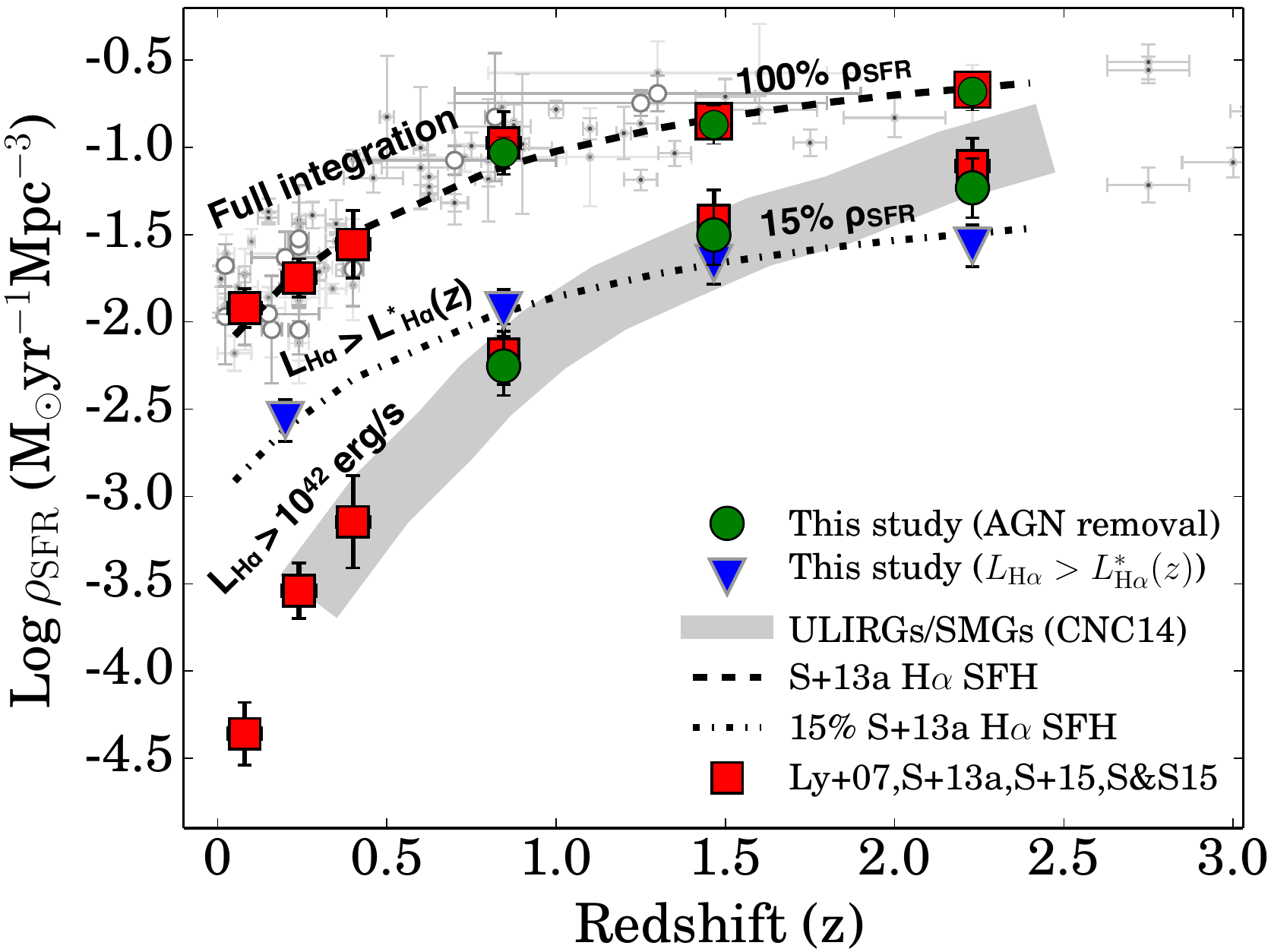}
\caption{The star formation rate density with increasing redshift. Our AGN-decontaminated star formation rate densities as applied to \citet{Sobral.13} are shown (green circles), and compared to the simpler fixed AGN contamination used in that paper of 10\% for $z<1$ \citep[][]{Garn.10} and 15\% for $z>1$ (red squares). The background data points for the full integration are from the literature \citep[e.g.][and references therein]{Hopkins.06}. While AGN-decontamination does not change the total integration of the $\rho_{\rm SFR}$ history noticeably (due to the dominant role of the $L_{\rm H\alpha}<L^*_{\rm H\alpha}$ sources at all redshifts), it becomes much more important for the most luminous sources, due to the higher AGN fractions at high luminosities. We find that the most luminous sources ($L_{\rm H\alpha}>L^*_{\rm H\alpha}$) exhibit an evolution in $\rho_{\rm SFR}$ that changes by $\sim$3 orders of magnitude over the redshifts shown when integrating down to $L_{\rm H\alpha}=10^{42}$\,erg\,s$^{-1}$ (observed, thus $L_{\rm H\alpha}=10^{42.4}$\,erg/\,s$^{-1}$ after dust corrected), which corresponds to $L_{\rm H\alpha}>3.4$\,$L^*_{\rm H\alpha}(z=0)$, but to only 0.34$L^*$ at $z=2.23$ (thus one is integrating further down the luminosity function as a function of increasing redshift). Such evolution is very well matched with that found for `ULIRGs' (defined in respect to $z=0$ with a fixed luminosity cut) and SMGs \citep[CNC14:][]{Casey2015}. However, if we take into account the evolution of $L^*_{\rm H\alpha}(z)$, and integrate the H$\alpha$ LF only above $L^*(z)$ at each redshift, and remove all AGN, we find that the fraction contribution of the extreme H$\alpha$ emitters is surprisingly constant across cosmic time, approximately $\sim15$\%. This reveals how it is misleading to integrate down to a fixed limit when luminosity functions are evolving very strongly in luminosity (thus the typical luminosity is changing).}
\label{fig:SFH}
\end{figure*}

Since we see that the AGN fraction is very high for H$\alpha$ emitters higher than $L^*$ at all epochs and $L^*$ is evolving very strongly with cosmic time, it is possible that the two are somewhat connected. However, this does not necessarily mean that AGN are quenching star formation. Indeed, we may just be witnessing that with more gas (and higher gas fractions), there is simply more accretion into the black hole (and more stars being formed) that is just driven by the gas supply without the AGN necessarily coupling to the SF \citep[e.g.][]{Mullaney2012}.

Even though our samples at each redshift are not very large, we also investigate if there is any strong evolution of the AGN fraction with redshift. Given that we find that the AGN fraction correlates very strongly with $L/L_{\rm H\alpha}^*(z)$, we take into account the $L/L_{\rm H\alpha}^*(z)$ distribution of the samples at the different redshifts ($z=0.8$, $z=1.47$, $z=2.23$). Our $z=0.8$ sample probes $L/L_{\rm H\alpha}^*(z)\sim1-6$ (average of 3.1), while we probe $L/L_{\rm H\alpha}^*(z)\sim1-50$ (average of 8) at $z=1.47$ and $L/L_{\rm H\alpha}^*(z)\sim1-23$ (average of 4) at z=2.23. This would imply, under the scenario of no AGN evolution with redshift, AGN fractions of $\sim20$\,\%, $\sim50$\,\% and $\sim30$\,\% at $z=0.8$, $z=1.47$ and $z=2.23$, respectively, while we find $6\pm6$\,\%, $50\pm16$\,\% and $21\pm14$\,\%. Thus, our results are consistent with no significant evolution of the AGN fraction with redshift, although there may be a slight decrease (at 2\,$\sigma$ significance) from $z\sim1.5-2.2$ to $z=0.8$. Larger samples at each individual redshifts would be required to further test this.

%
%
\begin{table*}
\caption{Results for our different samples. Note that the NB$_J$ sample is selected with two different narrow-band filters (see Figure \ref{FILTER_PROFILES}). $\phi$ (BL-AGN) is the number density of broad-line AGN. Note that the samples at $z\sim1.5$ and $z\sim2.2$ present a much larger AGN fraction and a much larger BL-AGN fraction, but they also reach much higher luminosities and higher L$^*$.}
\begin{tabular}{cccccccc}
\hline
Sample	&	$\bar{z}_{spec}$			&	Lookback time 	&	f$_{\rm BL-AGN}$	&  $\phi$ (BL-AGN) &	f$_{\rm AGN}$	&	obs. $\log$L$_{\rm H_{\alpha}}$   & L$_{\rm H\alpha}$/L$^*(z)$ \\	
	&				& [Gyr] 	&	 &  [$\times10^{-6}$\,Mpc$^{-3}$]	&	&	[$\log$(erg s$^{-1}$)]	 &  [L$^*(z)$]	\\
\hline						
NB$_{\rm J}$	& 0.84	& 7.0	& $6\pm6$  & $3\pm3$  & $6\pm6$ & 42.34 $\pm$ 0.18  & 1.2--6  \\
NB$_{\rm H}$	& 1.47	& 9.2	& $36\pm13$	& $9\pm3$   & $50\pm16$ & 43.01	$\pm$	0.35  & 1.9--50 \\
NB$_{\rm K}$	& 2.23	& 10.6	& $21\pm14$ 	& $3\pm2$ & $21\pm14$ & 43.16 $\pm$ 0.32 & 1.0--23 \\
\hline
\end{tabular}
\label{tab:averages}
\end{table*}

\subsection{AGN (de)contamination and an improvement on the accuracy of star-forming history among luminous H$\alpha$ emitters} \label{sub:SFH}

By removing AGN from our sample of luminous H$\alpha$ emitters, we derive the star formation rate density for such luminous sources and study its evolution. We present our results in Figure~\ref{fig:SFH} (green circles). We show the full integration of H$\alpha$ star formation rate density ($\rho_{\rm SFR}$) against redshift, with our AGN decontaminations applied to the three HiZELS redshift bins from \cite{Sobral.13}. We note that for all cases we use A$_{\rm H\alpha}=1$ for dust corrections \citep[see e.g.][]{Sobral.12,Ibar.13,Sobral.13}.

We present three different ways of investigating the evolution of the star formation rate density in Figure~\ref{fig:SFH}. The full integration presents the full star formation rate density in which the AGN decontamination at the bright end of the H$\alpha$ luminosity function has little effect. This reveals that even though the highest H$\alpha$ luminosity samples are significantly affected by AGN, the overall measurement is not affected significantly, because the star formation rate density at any epoch is dominated by the faintest H$\alpha$ emitters, for which the AGN fraction is low ($\sim10-15$\% at most).

We also present the star formation history when integrating down to roughly the combined H$\alpha$ luminosity limit of our samples, $L_{\rm H\alpha}=10^{42}$\,erg\,s$^{-1}$ (observed, thus $L_{\rm H\alpha}=10^{42.4}$\,erg\,s$^{-1}$ after dust correction), before and after removing AGN. Our results reveal the strong evolution of luminous H$\alpha$ star-forming galaxies across redshift, as $\rho_{\rm SFR}$ (down to a fixed luminosity limit) increases by a factor of $\sim1300\times$ over the range of redshifts shown, attributed to the most strongly star-forming H$\alpha$ selected galaxies from $z=0$ to $z=2.23$. Here the effect of AGN decontamination is much more important. Such rise in the contribution of highly star-forming systems to the total star-formation rate density, and the much stronger evolution with cosmic time, is also seen in extremely star-forming populations such as sub-millimetre galaxies (SMGs) or FIR selected galaxies at high flux thresholds, e.g. ULIRGs \citep[e.g.][]{Smail1997,Chapman2005,Caputi2007,Magnelli2011,Magnelli2013}.

Finally, we present the star formation rate density due to $L_{\rm H\alpha}>L_{\rm H\alpha}^*(z)$ star-forming galaxies (after removing all AGN). We use our samples for $z=0.8-2.23$, and also include the results at $z\sim0.2$ presented in \cite{Stroe.Sobral15} (assuming that the relation between AGN contamination and e.g. $L_{\rm H\alpha}/L_{\rm H\alpha}^*(z)$ does not evolve down to $z=0.2$), which provides a comparable narrow-band survey that successfully probes beyond L$^*$ and overcomes cosmic variance. Integrating down to $L>L^*(z)$ is a much fairer quantification of how much star formation density is occurring in the most star-forming galaxies at each redshift, as it takes into account that the typical H$\alpha$ luminosity (typical star formation rate; see e.g. \citealt{Sobral.14}) of galaxies is increasing with redshift. Once this is computed, the results shown in Figure ~\ref{fig:SFH} clearly reveal a very flat relative contribution of the most star-forming galaxies to the total star formation rate density, after removing AGN. This contribution is at the level of $\sim15$\,\%, independent of cosmic time. We note that such contribution matches very well the contribution of mergers to the total star-formation rate density \citep[e.g.][]{Sobral.09a,Stott13a}. Mergers have been found to dominate the H$\alpha$ luminosity function above $L_{\rm H\alpha}^*(z)$, at least at $z=0.8$ \citep[see][]{Sobral.09a}, and our results are consistent with this being the case at least up to $z\sim2.23$.

\section{Conclusions} \label{section:conc}

We have investigated the nature and evolution of the most luminous H$\alpha$ emitters across the peak of the star-formation and active galactic nuclei (AGN) activity in the Universe ($z=0.8-2.23$) by conducting spectroscopic observations with NTT/SofI, WHT/LIRIS and TNG/NICS. We selected 59 luminous H$\alpha$ emitters over three redshift slices ($>$ L$_{\rm H\alpha}^*$ at each epoch) at $z\sim0.8$, 1.5 and 2.2 from the HiZELS and CF-HiZELS surveys and obtained near-infrared spectra of these sources. By analysing their near-IR spectra we have unveiled their nature. Our main results are:

\begin{itemize}

\item We find that, overall, $30\pm8$\,\% of luminous H$\alpha$ emitters are AGN without any strong evolution with $z$ within the errors and particularly when taking into account the different H$\alpha$ luminosities probed). We find that $80\pm30$\% of the AGN among luminous H$\alpha$ emitters are broad-line AGN.

\item Our BL-AGN have black hole masses which span a relatively large range: from relatively typical black hole masses of a few $10^{7}$\,M$_{\odot}$ to more quasar like black hole masses at $z\sim2$ ($\sim10^{9}$\,M$_{\odot}$). These completely dominate the most luminous end of the H$\alpha$ luminosity function.

\item The AGN fraction and the fraction of broad-line AGN among luminous H$\alpha$ emitters increases strongly with H$\alpha$ flux, with H$\alpha$ luminosity and with $L/L^*(z)$ at all redshifts, with $L/L^*(z)$ being the strongest predictor of the AGN fraction and matching well the lower AGN fractions found for lower luminosity H$\alpha$ emitters.

\item While we find that $L_{\rm H\alpha}^*$ and lower luminosity H$\alpha$ emitters are dominated by star-forming galaxies, the most luminous H$\alpha$ emitters becoming increasingly AGN dominated at all cosmic epochs probed. $L>10L_{\rm H\alpha}^*(z)$) at any cosmic time are essentially all ($\sim100$\%) BL-AGN.

\item Using our AGN-decontaminated sample of star-forming galaxies, we also derive the star-formation history for the most luminous H$\alpha$ emitters since $z\sim2.23$. Our results reveal a factor of $\sim1300\times$ evolution in the star formation rate density attributed to the most strongly star-forming H$\alpha$ selected galaxies from $z=0$ to $z=2.23$. However, by integrating down to the evolving L$^*_{\rm H\alpha}(z)$, and classifying those as the most star-forming galaxies at any specific cosmic time, we show that the most star-forming galaxies at all redshifts up to $z\sim2.23$ have a constant contribution to the total star formation rate density of about 15\,\%. %

\end{itemize}

Our results are important in order to understand the nature and evolution of luminous H$\alpha$ emitters. We also find that the more luminous in H$\alpha$ a source is, the more likely it is to be an AGN, and the more likely it is to be a broad-line AGN, indicating that for the highest luminosities at any cosmic epoch, AGNs are the main powering mechanism. However, once one looks at more typical sources, the AGN fraction quickly reduces to $\sim10-15$\%.

\section*{Acknowledgments}

The authors would like to thank the anonymous reviewer for the many helpful comments and suggestions which greatly improved the clarity and quality of this work. D.S. and S.A.K. acknowledge financial support from the Netherlands Organisation for Scientific research (NWO) through a Veni fellowship. D.S. also acknowledges funding from FCT through a FCT Investigator Starting Grant and Start-up Grant (IF/01154/2012/CP0189/CT0010) and from FCT grant PEst-OE/FIS/UI2751/2014. Part of this project was undertaken during the inaugural Leiden/ESA Astrophysics Program for Summer Students (LEAPS). I.R.S. acknowledges support from STFC (ST/L00075X/1), the ERC Advanced Investigator programme DUSTYGAL 321334 and a Royal Society/Wolfson merit award. C.H. acknowledges support from STFC. Based on observations made with ESO Telescopes at the La Silla Paranal Observatory under programme ID 087.A-0337 and ID 089.A-0965. Also based on data from the Telescopio Nazionale Galileo, with time awarded through OPTICON programs 2011A/026 and 2012A020 and the William Herschel Telescope under program W12BN007. The William Herschel Telescope is operated on the island of La Palma by the Isaac Newton Group in the Spanish Observatorio del Roque de los Muchachos of the Instituto de Astrofisica de Canarias. The authors wish to thank all the help given by the telescope staff from all the observatories used in this study: ESO staff in La Silla, and the TNG and WHT staff in La Palma. This publication makes use of data products from the Two Micron All Sky Survey, which is a joint project of the University of Massachusetts and the Infrared Processing and Analysis Center/California Institute of Technology, funded by the National Aeronautics and Space Administration and the National Science Foundation.
  
\bibliographystyle{mn2e}
\bibliography{myBib}

\begin{thebibliography}{}

\bibitem[\protect\citeauthoryear{{Ackermann}, {Ajello}, {Allafort}, {Antolini},
  {Atwood}, {Axelsson}, {Baldini}, {Ballet} et~al.,}{{Ackermann}
  et~al.}{2011}]{Ackermann.11}
{Ackermann} M.,  {Ajello} M.,  {Allafort} A.,  {Antolini} E.,  {Atwood} W.~B.,
  {Axelsson} M.,  {Baldini} L.,  {Ballet} J.,    et~al., 2011, ApJ, 743, 171

\bibitem[\protect\citeauthoryear{{Aird}, {Nandra}, {Laird}, {Georgakakis},
  {Ashby}, {Barmby}, {Coil}, {Huang}, {Koekemoer}, {Steidel} \&
  {Willmer}}{{Aird} et~al.}{2010}]{Aird2010}
{Aird} J.,  {Nandra} K.,  {Laird} E.~S.,  {Georgakakis} A.,  {Ashby} M.~L.~N.,
  {Barmby} P.,  {Coil} A.~L.,  {Huang} J.-S.,  {Koekemoer} A.~M.,  {Steidel}
  C.~C.,    {Willmer} C.~N.~A.,  2010, MNRAS, 401, 2531

\bibitem[\protect\citeauthoryear{{An}, {Zheng}, {Wang} et~al.,}{{An}
  et~al.}{2014}]{An.14}
{An} F.~X.,  {Zheng} X.~Z.,  {Wang} W.-H.,    et~al., 2014, ApJ, 784, 152

\bibitem[\protect\citeauthoryear{{Baffa}, {Comoretto}, {Gennari}, {Lisi},
  {Oliva}, {Biliotti}, {Checcucci}, {Gavrioussev} et~al.,}{{Baffa}
  et~al.}{2001}]{Baffa.01}
{Baffa} C.,  {Comoretto} G.,  {Gennari} S.,  {Lisi} F.,  {Oliva} E.,
  {Biliotti} V.,  {Checcucci} A.,  {Gavrioussev} V.,    et~al., 2001, A\&A,
  378, 722

\bibitem[\protect\citeauthoryear{{Baldwin}, {Phillips} \&
  {Terlevich}}{{Baldwin} et~al.}{1981}]{Baldwin.81}
{Baldwin} J.~A.,  {Phillips} M.~M.,    {Terlevich} R.,  1981, PASP, 93, 5

\bibitem[\protect\citeauthoryear{{Best}, {Smail}, {Sobral} et~al.,}{{Best}
  et~al.}{2013}]{Best2013}
{Best} P.,  {Smail} I.,  {Sobral} D.,    et~al., 2013, ASSP, 37, 235

\bibitem[\protect\citeauthoryear{{Brand} et~al.,}{{Brand}
  et~al.}{2006}]{Brand.06}
{Brand} K.,  et~al., 2006, ApJ, 641, 140

\bibitem[\protect\citeauthoryear{{Brandt} \& {Alexander}}{{Brandt} \&
  {Alexander}}{2015}]{BrandtAlex15}
{Brandt} W.~N.,  {Alexander} D.~M.,  2015, AAPR, 23, 1

\bibitem[\protect\citeauthoryear{{Brinchmann}, {Charlot}, {White}, {Tremonti},
  {Kauffmann}, {Heckman} \& {Brinkmann}}{{Brinchmann}
  et~al.}{2004}]{Brinchmann.04}
{Brinchmann} J.,  {Charlot} S.,  {White} S.~D.~M.,  {Tremonti} C.,  {Kauffmann}
  G.,  {Heckman} T.,    {Brinkmann} J.,  2004, MNRAS, 351, 1151

\bibitem[\protect\citeauthoryear{{Calhau}, {Sobral} et~al.,}{{Calhau}
  et~al.}{2015}]{Calhau15}
{Calhau} J.,  {Sobral} D.,    et~al., 2015, MNRAS, submitted

\bibitem[\protect\citeauthoryear{{Caputi}, {Lagache}, {Yan}, {Dole},
  {Bavouzet}, {Le Floc'h}, {Choi}, {Helou} \& {Reddy}}{{Caputi}
  et~al.}{2007}]{Caputi2007}
{Caputi} K.~I.,  {Lagache} G.,  {Yan} L.,  {Dole} H.,  {Bavouzet} N.,  {Le
  Floc'h} E.,  {Choi} P.~I.,  {Helou} G.,    {Reddy} N.,  2007, ApJ, 660, 97

\bibitem[\protect\citeauthoryear{{Casey}, {Narayanan} \& {Cooray}}{{Casey}
  et~al.}{2014}]{Casey2015}
{Casey} C.~M.,  {Narayanan} D.,    {Cooray} A.,  2014, PhysREP, 541, 45

\bibitem[\protect\citeauthoryear{{Chapman}, {Blain}, {Smail} \&
  {Ivison}}{{Chapman} et~al.}{2005}]{Chapman2005}
{Chapman} S.~C.,  {Blain} A.~W.,  {Smail} I.,    {Ivison} R.~J.,  2005, ApJ,
  622, 772

\bibitem[\protect\citeauthoryear{{Colbert} et~al.,}{{Colbert}
  et~al.}{2013}]{Colbert.13}
{Colbert} J.~W.,  et~al., 2013, ApJ, 779, 34

\bibitem[\protect\citeauthoryear{{Darvish}, {Sobral}, {Mobasher}, {Scoville},
  {Best}, {Sales} \& {Smail}}{{Darvish} et~al.}{2014}]{Darvish2014}
{Darvish} B.,  {Sobral} D.,  {Mobasher} B.,  {Scoville} N.~Z.,  {Best} P.,
  {Sales} L.~V.,    {Smail} I.,  2014, ApJ, 796, 51

\bibitem[\protect\citeauthoryear{{Dom{\'{\i}}nguez} et~al.,}{{Dom{\'{\i}}nguez}
   et~al.}{2013}]{Dominguez.13}
{Dom{\'{\i}}nguez} A.,  et~al., 2013, ApJ, 763, 145

\bibitem[\protect\citeauthoryear{{Drake}, {Simpson}, {Baldry} et~al.,}{{Drake}
  et~al.}{2015}]{Drake2015}
{Drake} A.~B.,  {Simpson} C.,  {Baldry} I.~K.,    et~al., 2015,
  arXiv:1509.06900

\bibitem[\protect\citeauthoryear{{Fu}, {Yan}, {Scoville}, {Capak}, {Aussel},
  {Le Floc'h}, {Ilbert}, {Salvato} et~al.,}{{Fu} et~al.}{2010}]{Fu.10}
{Fu} H.,  {Yan} L.,  {Scoville} N.~Z.,  {Capak} P.,  {Aussel} H.,  {Le Floc'h}
  E.,  {Ilbert} O.,  {Salvato} M.,    et~al., 2010, ApJ, 722, 653

\bibitem[\protect\citeauthoryear{{Fumagalli} et~al.,}{{Fumagalli}
  et~al.}{2012}]{Fumagalli.12}
{Fumagalli} M.,  et~al., 2012, ApJL, 757, L22

\bibitem[\protect\citeauthoryear{{Garn} et~al.,}{{Garn}
  et~al.}{2010}]{Garn.10}
{Garn} T.,  et~al., 2010, MNRAS, 402, 2017

\bibitem[\protect\citeauthoryear{{Geach}, {Smail}, {Best}, {Kurk}, {Casali},
  {Ivison} \& {Coppin}}{{Geach} et~al.}{2008}]{Geach.08}
{Geach} J.~E.,  {Smail} I.,  {Best} P.~N.,  {Kurk} J.,  {Casali} M.,  {Ivison}
  R.~J.,    {Coppin} K.,  2008, MNRAS, 388, 1473

\bibitem[\protect\citeauthoryear{{Geach}, {Sobral}, {Hickox}, {Wake}, {Smail},
  {Best}, {Baugh} \& {Stott}}{{Geach} et~al.}{2012}]{Geach.12}
{Geach} J.~E.,  {Sobral} D.,  {Hickox} R.~C.,  {Wake} D.~A.,  {Smail} I.,
  {Best} P.~N.,  {Baugh} C.~M.,    {Stott} J.~P.,  2012, MNRAS, 426, 679

\bibitem[\protect\citeauthoryear{{Genzel}, {F{\"o}rster Schreiber}, {Rosario}
  et~al.,}{{Genzel} et~al.}{2014}]{Genzel2014}
{Genzel} R.,  {F{\"o}rster Schreiber} N.~M.,  {Rosario} D.,    et~al., 2014,
  ApJ, 796, 7

\bibitem[\protect\citeauthoryear{{Greene} \& {Ho}}{{Greene} \&
  {Ho}}{2005}]{Greene.05}
{Greene} J.~E.,  {Ho} L.~C.,  2005, ApJ, 630, 122

\bibitem[\protect\citeauthoryear{{Hayes}, {Schaerer} \& {{\"O}stlin}}{{Hayes}
  et~al.}{2010}]{Hayes2010}
{Hayes} M.,  {Schaerer} D.,    {{\"O}stlin} G.,  2010, A\&A, 509, L5

\bibitem[\protect\citeauthoryear{{Heckman}}{{Heckman}}{2003}]{Heckman.03}
{Heckman} T.~M.,  2003 Vol.~17 of Revista Mexicana de Astronomia y Astrofisica
  Conference Series, {Starburst-Driven Galactic Winds}.
pp 47--55

\bibitem[\protect\citeauthoryear{{Heckman} \& {Best}}{{Heckman} \&
  {Best}}{2014}]{HeckBest2014}
{Heckman} T.~M.,  {Best} P.~N.,  2014, ARAA, 52, 589

\bibitem[\protect\citeauthoryear{{Ho}, {Filippenko}, {Sargent} \& {Peng}}{{Ho}
  et~al.}{1997}]{Ho.97}
{Ho} L.~C.,  {Filippenko} A.~V.,  {Sargent} W.~L.~W.,    {Peng} C.~Y.,  1997,
  ApJS, 112, 391

\bibitem[\protect\citeauthoryear{{Hopkins} \& {Beacom}}{{Hopkins} \&
  {Beacom}}{2006}]{Hopkins.06}
{Hopkins} A.~M.,  {Beacom} J.~F.,  2006, ApJ, 651, 142

\bibitem[\protect\citeauthoryear{{Ibar} et~al.,}{{Ibar}
  et~al.}{2013}]{Ibar.13}
{Ibar} E.,  et~al., 2013, MNRAS, 434, 3218

\bibitem[\protect\citeauthoryear{{Karim} et~al.,}{{Karim}
  et~al.}{2011}]{Karim.11}
{Karim} A.,  et~al., 2011, ApJ, 730, 61

\bibitem[\protect\citeauthoryear{{Kennicutt}
  Jr.}{{Kennicutt}}{1998}]{Kennicutt.98}
{Kennicutt} Jr. R.~C.,  1998, ARA\&A, 36, 189

\bibitem[\protect\citeauthoryear{{Kewley}, {Maier}, {Yabe}, {Ohta}, {Akiyama},
  {Dopita} \& {Yuan}}{{Kewley} et~al.}{2013}]{Kewley.13}
{Kewley} L.~J.,  {Maier} C.,  {Yabe} K.,  {Ohta} K.,  {Akiyama} M.,  {Dopita}
  M.~A.,    {Yuan} T.,  2013, ApJL, 774, L10

\bibitem[\protect\citeauthoryear{{Khostovan}, {Sobral}, {Mobasher}, {Best},
  {Smail}, {Stott}, {Hemmati} \& {Nayyeri}}{{Khostovan}
  et~al.}{2015}]{Khostovan2015}
{Khostovan} A.~A.,  {Sobral} D.,  {Mobasher} B.,  {Best} P.~N.,  {Smail} I.,
  {Stott} J.~P.,  {Hemmati} S.,    {Nayyeri} H.,  2015, MNRAS, 452, 3948

\bibitem[\protect\citeauthoryear{{Kirkpatrick}, {Pope}, {Alexander},
  {Charmandaris}, {Daddi}, {Dickinson}, {Elbaz}, {Gabor} et~al.,}{{Kirkpatrick}
  et~al.}{2012}]{Kirkpatrick.12}
{Kirkpatrick} A.,  {Pope} A.,  {Alexander} D.~M.,  {Charmandaris} V.,  {Daddi}
  E.,  {Dickinson} M.,  {Elbaz} D.,  {Gabor} J.,    et~al., 2012, ApJ, 759, 139

\bibitem[\protect\citeauthoryear{{Koyama} et~al.,}{{Koyama}
  et~al.}{2013}]{Koyama.13}
{Koyama} Y.,  et~al., 2013, MNRAS, 434, 423

\bibitem[\protect\citeauthoryear{{Kriek} et~al.,}{{Kriek}
  et~al.}{2014}]{Kriek15}
{Kriek} M.,  et~al., 2014, arXiv:1412.1835

\bibitem[\protect\citeauthoryear{{Kurk}, {Pentericci}, {Overzier},
  {R{\"o}ttgering} \& {Miley}}{{Kurk} et~al.}{2004}]{Kurk2004}
{Kurk} J.~D.,  {Pentericci} L.,  {Overzier} R.~A.,  {R{\"o}ttgering} H.~J.~A.,
    {Miley} G.~K.,  2004, A\&A, 428, 817

\bibitem[\protect\citeauthoryear{{Lacy} et~al.,}{{Lacy}
  et~al.}{2004}]{Lacy.04}
{Lacy} M.,  et~al., 2004, ApJS, 154, 166

\bibitem[\protect\citeauthoryear{{Lacy} et~al.,}{{Lacy}
  et~al.}{2007}]{Lacy.07}
{Lacy} M.,  et~al., 2007, AJ, 133, 186

\bibitem[\protect\citeauthoryear{{LaMassa}, {Heckman}, {Ptak}, {Schiminovich},
  {O'Dowd} \& {Bertincourt}}{{LaMassa} et~al.}{2012}]{LaMassa.12}
{LaMassa} S.~M.,  {Heckman} T.~M.,  {Ptak} A.,  {Schiminovich} D.,  {O'Dowd}
  M.,    {Bertincourt} B.,  2012, ApJ, 758, 1

\bibitem[\protect\citeauthoryear{{Lawrence}, {Warren}, {Almaini}, {Edge},
  {Hambly}, {Jameson}, {Lucas}, {Casali} et~al.,}{{Lawrence}
  et~al.}{2012}]{Lawrence.12}
{Lawrence} A.,  {Warren} S.~J.,  {Almaini} O.,  {Edge} A.~C.,  {Hambly} N.~C.,
  {Jameson} R.~F.,  {Lucas} P.,  {Casali} M.,    et~al., 2012, VizieR Online
  Data Catalog, 2314, 0

\bibitem[\protect\citeauthoryear{{Lee}, {Ly}, {Spitler}, {Labb{\'e}}, {Salim},
  {Persson}, {Ouchi}, {Dale}, {Monson} \& {Murphy}}{{Lee}
  et~al.}{2012}]{Lee2012}
{Lee} J.~C.,  {Ly} C.,  {Spitler} L.,  {Labb{\'e}} I.,  {Salim} S.,  {Persson}
  S.~E.,  {Ouchi} M.,  {Dale} D.~A.,  {Monson} A.,    {Murphy} D.,  2012, PASP,
  124, 782

\bibitem[\protect\citeauthoryear{{Lilly}, {Le Fevre}, {Hammer} \&
  {Crampton}}{{Lilly} et~al.}{1996}]{Lilly.96}
{Lilly} S.~J.,  {Le Fevre} O.,  {Hammer} F.,    {Crampton} D.,  1996, ApJL,
  460, L1

\bibitem[\protect\citeauthoryear{{Livermore} et~al.,}{{Livermore}
  et~al.}{2012}]{Livermore.12}
{Livermore} R.~C.,  et~al., 2012, MNRAS, 427, 688

\bibitem[\protect\citeauthoryear{{Magnelli}, {Elbaz}, {Chary}, {Dickinson}, {Le
  Borgne}, {Frayer} \& {Willmer}}{{Magnelli} et~al.}{2011}]{Magnelli2011}
{Magnelli} B.,  {Elbaz} D.,  {Chary} R.~R.,  {Dickinson} M.,  {Le Borgne} D.,
  {Frayer} D.~T.,    {Willmer} C.~N.~A.,  2011, A\&A, 528, A35

\bibitem[\protect\citeauthoryear{{Magnelli}, {Popesso}, {Berta}
  et~al.,}{{Magnelli} et~al.}{2013}]{Magnelli2013}
{Magnelli} B.,  {Popesso} P.,  {Berta} S.,    et~al., 2013, A\&A, 553, A132

\bibitem[\protect\citeauthoryear{{Manchado}, {Fuentes}, {Prada}, {Ballesteros},
  {Barreto}, {Carranza}, {Escudero}, {Fragoso-Lopez} et~al.,}{{Manchado}
  et~al.}{1998}]{Manchado.98}
{Manchado} A.,  {Fuentes} F.~J.,  {Prada} F.,  {Ballesteros} E.,  {Barreto} M.,
   {Carranza} J.~M.,  {Escudero} I.,  {Fragoso-Lopez} A.~B.,    et~al., 1998
  Vol.~3354 of Society of Photo-Optical Instrumentation Engineers (SPIE)
  Conference Series.
pp 448--455

\bibitem[\protect\citeauthoryear{{Matthee}, {Sobral}, {Swinbank}
  et~al.,}{{Matthee} et~al.}{2014}]{Matthee2014}
{Matthee} J.~J.~A.,  {Sobral} D.,  {Swinbank} A.~M.,    et~al., 2014, MNRAS,
  440, 2375

\bibitem[\protect\citeauthoryear{{McLean}, {Steidel}, {Matthews}, {Epps} \&
  {Adkins}}{{McLean} et~al.}{2008}]{McLean.08}
{McLean} I.~S.,  {Steidel} C.~C.,  {Matthews} K.,  {Epps} H.,    {Adkins}
  S.~M.,  2008 Vol.~7014 of Society of Photo-Optical Instrumentation Engineers
  (SPIE) Conference Series.
p. 12pp.

\bibitem[\protect\citeauthoryear{{McLure} \& {Dunlop}}{{McLure} \&
  {Dunlop}}{2004}]{McLure04}
{McLure} R.~J.,  {Dunlop} J.~S.,  2004, MNRAS, 352, 1390

\bibitem[\protect\citeauthoryear{{Moorwood}, {Cuby} \& {Lidman}}{{Moorwood}
  et~al.}{1998}]{Moorwood.98}
{Moorwood} A.,  {Cuby} J.-G.,    {Lidman} C.,  1998, The Messenger, 91, 9

\bibitem[\protect\citeauthoryear{{Mullaney}, {Daddi}, {B{\'e}thermin}, {Elbaz},
  {Juneau}, {Pannella}, {Sargent}, {Alexander} \& {Hickox}}{{Mullaney}
  et~al.}{2012}]{Mullaney2012}
{Mullaney} J.~R.,  {Daddi} E.,  {B{\'e}thermin} M.,  {Elbaz} D.,  {Juneau} S.,
  {Pannella} M.,  {Sargent} M.~T.,  {Alexander} D.~M.,    {Hickox} R.~C.,
  2012, ApJL, 753, L30

\bibitem[\protect\citeauthoryear{{Newman}, {Genzel}, {F{\"o}rster-Schreiber}
  et~al.,}{{Newman} et~al.}{2012}]{Newman2012}
{Newman} S.~F.,  {Genzel} R.,  {F{\"o}rster-Schreiber} N.~M.,    et~al., 2012,
  ApJ, 761, 43

\bibitem[\protect\citeauthoryear{{Obri{\'c}}, {Ivezi{\'c}}, {Best}, {Lupton},
  {Tremonti}, {Brinchmann}, {Ag{\"u}eros}, {Knapp} et~al.,}{{Obri{\'c}}
  et~al.}{2006}]{Obric.06}
{Obri{\'c}} M.,  {Ivezi{\'c}} {\v Z}.,  {Best} P.~N.,  {Lupton} R.~H.,
  {Tremonti} C.,  {Brinchmann} J.,  {Ag{\"u}eros} M.~A.,  {Knapp} G.~R.,
  et~al., 2006, MNRAS, 370, 1677

\bibitem[\protect\citeauthoryear{{Osterbrock}, {Fulbright}, {Martel}, {Keane},
  {Trager} \& {Basri}}{{Osterbrock} et~al.}{1996}]{Osterbrock.96}
{Osterbrock} D.~E.,  {Fulbright} J.~P.,  {Martel} A.~R.,  {Keane} M.~J.,
  {Trager} S.~C.,    {Basri} G.,  1996, PASP, 108, 277

\bibitem[\protect\citeauthoryear{{Price}, {Kriek}, {Brammer} et~al.,}{{Price}
  et~al.}{2014}]{Price.13}
{Price} S.~H.,  {Kriek} M.,  {Brammer} G.~B.,    et~al., 2014, ApJ, 788, 86

\bibitem[\protect\citeauthoryear{{Reddy}, {Steidel}, {Pettini}, {Adelberger},
  {Shapley}, {Erb} \& {Dickinson}}{{Reddy} et~al.}{2008}]{Reddy.08}
{Reddy} N.~A.,  {Steidel} C.~C.,  {Pettini} M.,  {Adelberger} K.~L.,  {Shapley}
  A.~E.,  {Erb} D.~K.,    {Dickinson} M.,  2008, ApJS, 175, 48

\bibitem[\protect\citeauthoryear{{Rola}, {Terlevich} \& {Terlevich}}{{Rola}
  et~al.}{1997}]{Rola.97}
{Rola} C.~S.,  {Terlevich} E.,    {Terlevich} R.~J.,  1997, MNRAS, 289, 419

\bibitem[\protect\citeauthoryear{{Schenker}, {Ellis}, {Konidaris} \&
  {Stark}}{{Schenker} et~al.}{2013}]{Schenker.13}
{Schenker} M.~A.,  {Ellis} R.~S.,  {Konidaris} N.~P.,    {Stark} D.~P.,  2013,
  ApJ, 777, 67

\bibitem[\protect\citeauthoryear{{Scoville}, {Aussel}, {Brusa}, {Capak},
  {Carollo}, {Elvis}, {Giavalisco}, {Guzzo} et~al.,}{{Scoville}
  et~al.}{2007}]{Scoville.07}
{Scoville} N.,  {Aussel} H.,  {Brusa} M.,  {Capak} P.,  {Carollo} C.~M.,
  {Elvis} M.,  {Giavalisco} M.,  {Guzzo} L.,    et~al., 2007, ApJS, 172, 1

\bibitem[\protect\citeauthoryear{{Shapley}, {Reddy}, {Kriek} et~al.,}{{Shapley}
  et~al.}{2015}]{Shapley2015}
{Shapley} A.~E.,  {Reddy} N.~A.,  {Kriek} M.,    et~al., 2015, ApJ, 801, 88

\bibitem[\protect\citeauthoryear{{Sharples} et~al.,}{{Sharples}
  et~al.}{2006}]{Sharples.06}
{Sharples} R.,  et~al., 2006, NewAR, 50, 370

\bibitem[\protect\citeauthoryear{{Skrutskie}, {Cutri}, {Stiening}, {Weinberg},
  {Schneider}, {Carpenter}, {Beichman}, {Capps} et~al.,}{{Skrutskie}
  et~al.}{2006}]{Skrutskie.06}
{Skrutskie} M.~F.,  {Cutri} R.~M.,  {Stiening} R.,  {Weinberg} M.~D.,
  {Schneider} S.,  {Carpenter} J.~M.,  {Beichman} C.,  {Capps} R.,    et~al.,
  2006, AJ, 131, 1163

\bibitem[\protect\citeauthoryear{{Smail}, {Ivison} \& {Blain}}{{Smail}
  et~al.}{1997}]{Smail1997}
{Smail} I.,  {Ivison} R.~J.,    {Blain} A.~W.,  1997, ApJL, 490, L5

\bibitem[\protect\citeauthoryear{{Smol{\v c}i{\'c}}, {Ivezi{\'c}}, {Ga{\'c}e{\v
  s}a}, {Rakos}, {Pavlovski}, {Iliji{\'c}}, {Obri{\'c}}, {Lupton}
  et~al.,}{{Smol{\v c}i{\'c}} et~al.}{2006}]{Smolcic.06}
{Smol{\v c}i{\'c}} V.,  {Ivezi{\'c}} {\v Z}.,  {Ga{\'c}e{\v s}a} M.,  {Rakos}
  K.,  {Pavlovski} K.,  {Iliji{\'c}} S.,  {Obri{\'c}} M.,  {Lupton} R.~H.,
  et~al., 2006, MNRAS, 371, 121

\bibitem[\protect\citeauthoryear{{Sobral}, {Best}, {Geach}, {Smail},
  {Cirasuolo}, {Garn}, {Dalton} \& {Kurk}}{{Sobral} et~al.}{2010}]{Sobral.10}
{Sobral} D.,  {Best} P.~N.,  {Geach} J.~E.,  {Smail} I.,  {Cirasuolo} M.,
  {Garn} T.,  {Dalton} G.~B.,    {Kurk} J.,  2010, MNRAS, 404, 1551

\bibitem[\protect\citeauthoryear{{Sobral}, {Best}, {Geach}, {Smail}, {Kurk},
  {Cirasuolo}, {Casali}, {Ivison} et~al.,}{{Sobral} et~al.}{2009}]{Sobral.09a}
{Sobral} D.,  {Best} P.~N.,  {Geach} J.~E.,  {Smail} I.,  {Kurk} J.,
  {Cirasuolo} M.,  {Casali} M.,  {Ivison} R.~J.,    et~al., 2009, MNRAS, 398,
  75

\bibitem[\protect\citeauthoryear{{Sobral}, {Best}, {Matsuda}, {Smail}, {Geach}
  \& {Cirasuolo}}{{Sobral} et~al.}{2012}]{Sobral.12}
{Sobral} D.,  {Best} P.~N.,  {Matsuda} Y.,  {Smail} I.,  {Geach} J.~E.,
  {Cirasuolo} M.,  2012, MNRAS, 420, 1926

\bibitem[\protect\citeauthoryear{{Sobral}, {Best}, {Smail}, {Mobasher}, {Stott}
  \& {Nisbet}}{{Sobral} et~al.}{2014}]{Sobral.14}
{Sobral} D.,  {Best} P.~N.,  {Smail} I.,  {Mobasher} B.,  {Stott} J.,
  {Nisbet} D.,  2014, MNRAS, 437, 3516

\bibitem[\protect\citeauthoryear{{Sobral} et~al.,}{{Sobral}
  et~al.}{2013a}]{Sobral.13}
{Sobral} D.,  et~al., 2013a, MNRAS, 428, 1128

\bibitem[\protect\citeauthoryear{{Sobral} et~al.,}{{Sobral}
  et~al.}{2013b}]{Sobral.13b}
{Sobral} D.,  et~al., 2013b, MNRAS, 779, 139

\bibitem[\protect\citeauthoryear{{Sobral}, {Matthee}, {Best}, {Smail},
  {Khostovan}, {Milvang-Jensen}, {Kim}, {Stott}, {Calhau}, {Nayyeri} \&
  {Mobasher}}{{Sobral} et~al.}{2015}]{Sobral15}
{Sobral} D.,  {Matthee} J.,  {Best} P.~N.,  {Smail} I.,  {Khostovan} A.~A.,
  {Milvang-Jensen} B.,  {Kim} J.-W.,  {Stott} J.,  {Calhau} J.,  {Nayyeri} H.,
    {Mobasher} B.,  2015, MNRAS, 451, 2303

\bibitem[\protect\citeauthoryear{{Sobral}, {Stroe}, {Dawson}, {Wittman}, {Jee},
  {R{\"o}ttgering}, {van Weeren} \& {Br{\"u}ggen}}{{Sobral}
  et~al.}{2015}]{Sobral2015S}
{Sobral} D.,  {Stroe} A.,  {Dawson} W.~A.,  {Wittman} D.,  {Jee} M.~J.,
  {R{\"o}ttgering} H.,  {van Weeren} R.~J.,    {Br{\"u}ggen} M.,  2015, MNRAS,
  450, 630

\bibitem[\protect\citeauthoryear{{Stern} et~al.,}{{Stern}
  et~al.}{2012}]{Stern.12}
{Stern} D.,  et~al., 2012, ApJ, 753, 30

\bibitem[\protect\citeauthoryear{{Stirpe}}{{Stirpe}}{1990}]{Stirpe.90}
{Stirpe} G.~M.,  1990, A\&A, 85, 1049

\bibitem[\protect\citeauthoryear{{Stott}, {Sobral} et~al.,}{{Stott}
  et~al.}{2013a}]{Stott13a}
{Stott} J.~P.,  {Sobral} D.,    et~al., 2013a, MNRAS, 430, 1158

\bibitem[\protect\citeauthoryear{{Stott}, {Sobral} et~al.,}{{Stott}
  et~al.}{2013b}]{Stott13b}
{Stott} J.~P.,  {Sobral} D.,    et~al., 2013b, MNRAS, 436, 1130

\bibitem[\protect\citeauthoryear{{Stott}, {Sobral}, {Swinbank}, {Smail},
  {Bower}, {Best}, {Sharples}, {Geach} \& {Matthee}}{{Stott}
  et~al.}{2014}]{Stott2014}
{Stott} J.~P.,  {Sobral} D.,  {Swinbank} A.~M.,  {Smail} I.,  {Bower} R.,
  {Best} P.~N.,  {Sharples} R.~M.,  {Geach} J.~E.,    {Matthee} J.,  2014,
  MNRAS, 443, 2695

\bibitem[\protect\citeauthoryear{{Stroe} \& {Sobral}}{{Stroe} \&
  {Sobral}}{2015}]{Stroe.Sobral15}
{Stroe} A.,  {Sobral} D.,  2015, MNRAS, 453, 242

\bibitem[\protect\citeauthoryear{{Swinbank}, {Simpson}, {Smail}
  et~al.,}{{Swinbank} et~al.}{2014}]{Swinbank2014}
{Swinbank} A.~M.,  {Simpson} J.~M.,  {Smail} I.,    et~al., 2014, MNRAS, 438,
  1267

\bibitem[\protect\citeauthoryear{{Swinbank}, {Sobral}, {Smail}, {Geach},
  {Best}, {McCarthy}, {Crain} \& {Theuns}}{{Swinbank}
  et~al.}{2012}]{Swinbank.12}
{Swinbank} A.~M.,  {Sobral} D.,  {Smail} I.,  {Geach} J.~E.,  {Best} P.~N.,
  {McCarthy} I.~G.,  {Crain} R.~A.,    {Theuns} T.,  2012, MNRAS, 426, 935

\bibitem[\protect\citeauthoryear{{Tadaki}, {Kodama}, {Tanaka}, {Hayashi},
  {Koyama} \& {Shimakawa}}{{Tadaki} et~al.}{2013}]{Tadaki.13}
{Tadaki} K.-i.,  {Kodama} T.,  {Tanaka} I.,  {Hayashi} M.,  {Koyama} Y.,
  {Shimakawa} R.,  2013, ApJ, 778, 114

\bibitem[\protect\citeauthoryear{{Trump}, {Konidaris}, {Barro}, {Koo},
  {Kocevski}, {Juneau}, {Weiner}, {Faber} et~al.,}{{Trump}
  et~al.}{2013}]{Trump.13}
{Trump} J.~R.,  {Konidaris} N.~P.,  {Barro} G.,  {Koo} D.~C.,  {Kocevski}
  D.~D.,  {Juneau} S.,  {Weiner} B.~J.,  {Faber} S.~M.,    et~al., 2013, ApJL,
  763, L6

\bibitem[\protect\citeauthoryear{{Ueda}, {Akiyama}, {Hasinger}, {Miyaji} \&
  {Watson}}{{Ueda} et~al.}{2014}]{Ueda2014}
{Ueda} Y.,  {Akiyama} M.,  {Hasinger} G.,  {Miyaji} T.,    {Watson} M.~G.,
  2014, ApJ, 786, 104

\bibitem[\protect\citeauthoryear{{Wolf}, {Wisotzki}, {Borch}, {Dye},
  {Kleinheinrich} \& {Meisenheimer}}{{Wolf} et~al.}{2003}]{Wolf.03}
{Wolf} C.,  {Wisotzki} L.,  {Borch} A.,  {Dye} S.,  {Kleinheinrich} M.,
  {Meisenheimer} K.,  2003, A\&A, 408, 499

\bibitem[\protect\citeauthoryear{{Wuyts}, {Kurk}, {F{\"o}rster Schreiber}
  et~al.,}{{Wuyts} et~al.}{2014}]{Wuyts.14}
{Wuyts} E.,  {Kurk} J.,  {F{\"o}rster Schreiber} N.~M.,    et~al., 2014, ApJL,
  789, L40

\end{thebibliography}

\appendix

%
%
\begin{table*}
 \centering
  \caption{Example entries from the catalogue of 59 sources. The full catalogue is available on-line.}
  \begin{tabular}{@{}cccccccccccccc@{}}
  \hline
   ID & R.A. & Dec. & $z_{\rm spec}$ & log L$_{\rm H\alpha}$  & L$_{H\alpha}$/L$_{\rm H\alpha}^*$   & FWHM H$\alpha$ & [NII]/H$\alpha$   & [OIII]/H$\beta$  & Class & Instrum.  \\
         & {(J2000)} &(J2000) &  &erg\,s$^{-1}$ &  & km\,s$^{-1}$ & log & log &  \\
 \hline
   \noalign{\smallskip}
BR-03 & 02:19:08.8 & -04:40:35.7 & $1.4845\pm0.0005$ & 42.46 & 2.1 & $430\pm91$ & $-0.89\pm0.42$ & $0.54\pm0.30$ & SFG & SofI \\ 
BR-04 & 02:17:08.7 &  -04:57:41.5 & $1.4394\pm0.0009$ & 42.65 & 3.4 & $2225\pm168$ &  $-0.94\pm0.42$ & $-0.06\pm0.35$ & BL-AGN &  SofI \\
BR-05 & 02:17:37.2 & -04:46:12.3 & $1.4621\pm0.0002$ &  42.68 & 3.5 & $979\pm167$ &$-0.55\pm0.44$ & $0.26\pm0.36$ & SFG & SofI \\
 \hline
\end{tabular}
\label{NBJ_CAT}
\end{table*}

\label{lastpage}

\end{document}